\documentstyle[epsf]{elsart}

\date{\ }

\begin{document}
\begin{frontmatter}

\title{Supersymmetric Field Theory
of Non--Equilibrium Thermodynamic System}

\author{Alexander I. Olemskoi and Valerii A. Brazhnyi\thanksref{AOVB}}
\address{Physical Electronics Department, Sumy State University\\
2, Rimskii-Korsakov Str., 244007 Sumy UKRAINE}

\thanks[AOVB]{olemskoi\char'100ssu.sumy.ua,
valera\char'100ssu.sumy.ua}

\begin{abstract}

On the basis of Langevin equation the optimal SUSY field scheme is
formulated to discribe a non--equilibrium thermodynamic system with
quenched disorder and non--ergodicity effects.
Thermodynamic and isothermal susceptibilities, memory parameter and
irreversible response are determined at different
temperatures and quenched disorder intensities.

\vspace{0.5cm}

{\it PACS:} 05.70.Ln; 11.30.Pb; 64.60.--i

\vspace{0.5cm}

{\it Keywords:} {Order Parameter; Fluctuatuon; Quenched disorder;
Non--ergodicity Parameter}

\end{abstract}

\end{frontmatter}

\section  {Introduction }

Recently, the microscopic theory of non--equilibrium thermodynamic
systems with broken ergodicity and exhibiting the memory effect, has been
the subject of major interest.
Spin glasses \cite{1} and random heteropolymers \cite{2}, that have
received much consideration, are well-known examples of such
systems. Despite the bulk of  theoretical studies had been employed the
replica method to approach the problem analytically, there
is increasing interest in alternative methods that go beyond the
replica trick. The
supersymmetry (SUSY) approach, evolved within stochastic dynamics
theory governed by Langevin equation, gives
a good example of method of such kind.

According to this method,
generating functional of Langevin dynamical system is represented as
a functional integral over the superfields with Euclidean action by
means of introducing Grassmann anticommutating variables. These
variables and their products serve as a basis for superfields
with components that involve
Grassmann fields along with real(complex)--valued ones.

As it was shown in \cite{3},
static replica treatment of spin systems bears striking similarity
with the dynamics
expressed in terms of superspace within the framework of
the SUSY method.
The latter is based on
using of nilpotent variables.
Two--point correlator of SUSY field
can be written in the form of expansion
with coefficients that give correlators of
observables such as structure factor $ S $, and
retarded and advanced  Green functions $ G_\mp $.
The memory and nonergodicity effects
are allowed for by incorporating the additional
terms $ q $, $ \Delta $ into the correlators.
The resulting
self--consistent SUSY scheme gives a set of equations
for memory $ q $ and nonergodicity $ \Delta $ parameters
to be determined as functions
of temperature $T$ and quenched disorder $h$.

For SUSY scheme formulation SUSY as a gauge field needs to be
reduced to irreducible components. By analogy with electromagnetic field,
that can be splitted into vector and scalar fields, 4--component SUSY field
can be divided into chiral  components that consist of regular and Grassmann
constituents \cite{4}. In Sections 2 and 3 SUSY field will  be
reduced to  $2$--component nilpotent field.
The latter has an advantage over conventional SUSY representation
because its components
have an explicit physical meaning of order parameter and conjugate
field (or amplitude of its fluctuation).
This rises the question as to  optimal choice of the basis for making
expansion of SUSY
correlators. Currently, two types of such basis are known
\cite{5,6}. The first one contains 3 components: advanced and retarded
Green functions $G_\pm$ and structure factor $S$. The second basis
corresponds to the proper $4$--component SUSY field and contains $5$
components which, in addition to the above mentioned ones,
include a couple of mutually
conjugated correlators of the Grassmann fields. In Sect.4
it will be
shown that the second basis can be reduced to the first one.

The work is organized as follows.
In Sect.2, the simplest
field scheme is formulated  in terms of
the $2$--component nilpotent fields with second component
is taken to be
either an amplitude of fluctuation or a conjugate force (see
subsections 2.1 and 2.2).
In Sect.3 the
above--discussed method for reduction of the $4$--component proper
SUSY field to different $2$--component forms is presented.
In Sect.4 we show that
the reduction results in decrease of the number of components of the
SUSY correlator basis, due to the fact that the conjugate correlators
of the Grassmann fields are equal to the retarded Green function in
accordance with Ward identities. The SUSY perturbation theory, stated
for both cubic and quadratic anharmonicities in Sect.5, makes
expressions for the SUSY self--energy function simple to calculate.
This function enters  the SUSY Dyson equation derived
in Sect.6 on the basis of effective Lagrangians for both
thermodynamic systems with quenched disorder and random
heteropolymers.
Nonergodicity and memory effects are
investigated in Sect.7. The corresponding self--consistent
equations are obtained. Behavior of a non--equilibrium
thermodynamic system for various values of temperature and
quenched disorder intensity is analyzed in Sect.8.
Appendices A, B, C provide details concerning
the SUSY formalism under consideration.
\section {Two--component SUSY representation}

Let us start with the simplest stochastic Langevin equation \cite {7}
governing the spatiotemporal evolution of order parameter
$ \eta ({\bf r}, t) $:

\begin {equation}
\dot{\eta}({\bf r},t) - D\nabla^2\eta=
-\gamma (\partial V/\partial \eta) +\zeta({\bf r},t),
\label{1}
\end {equation}
where the dot stands for derivative with respect to time,
$\nabla\equiv\partial/\partial {\bf r}$,
$D$ is the
diffusion--like coefficient,
$\gamma$ is the kinetic coefficient, $V(\eta)$ is
the synergetic potential (Landau free energy),
$ \zeta ({\bf r}, t) $ is
a Gaussian stochastic function subjected to the white noise conditions

\begin{equation}
\langle\zeta({\bf r},t) \rangle_{0}=0, \qquad \left\langle
\zeta({\bf r},t)\zeta({\bf 0},0)\right\rangle_{0}=\gamma T\delta({\bf
r})\delta(t),
\label{2}
\end {equation}
where the angular brackets with subscript $0$ denote
averaging over the Gaussian probability distribution of $\zeta$,
$ T $ is the intensity
of the noise (the temperature of thermostat).

Further, it
is convenient to measure
time $t$, coordinate ${\bf r}$,
synergetic potential $V$, and stochastic variable $\zeta$,
in units $t_s\equiv (\gamma
T)^2/D^3$, $r_s\equiv \gamma T/D$, $V_s\equiv D^3/\gamma^3T^2$,
$\zeta_s\equiv D^3/(\gamma T)^2$
respectively. The equation of motion (\ref {1}) then reads

\begin{equation}
\dot{\eta}({\bf r},t) = -\delta V/\delta\eta + \zeta({\bf r},t),
\label{3}
\end {equation}
where short notation is used for the variational derivative

\begin{equation}
\delta V/\delta \eta\equiv
\delta V\{\eta\}/\delta\eta=\partial V(\eta)/\partial\eta
-\nabla^2\eta, \qquad
V\{\eta\}\equiv
\int \left[
V(\eta)+\frac{1}{2}
\left(\nabla\eta\right)^2
\right]\,
{\rm d}{\bf r},
\label{4}
\end {equation}
the
coefficient $\gamma T$ in Eq.(\ref{2}) becomes unity and  the
distribution of variable $\zeta$ takes the Gaussian form

\begin{equation}
P_0\{\zeta\}\propto \exp\left(- {1\over 2}\int \zeta^2({\bf
r},t){\rm d}{\bf r}{\rm d}t\right).
\label{5}
\end{equation}

The basis for construction of the field scheme is the generating
functional \cite {8}

\begin{eqnarray}
&&Z\{ u({\bf r},t) \} = \int
\!\!Z\{ \eta \} \exp \left( \int \!
u\eta \, {\rm d}{\bf r}\, {\rm d}t\right) {\rm
D}\eta , \label{6} \\
&&Z\{ \eta ({\bf r},t)\} \equiv \left< \prod_{({\bf
r},t)} \delta \left\{ \dot{\eta} +{\delta V\over
\delta\eta }-\zeta \right\} {\rm
det}\left|{\delta\zeta \over \delta\eta} \right|\right>_0,
\label{7}
\end{eqnarray}
so that its variational derivatives with respect to an auxiliary field
$ u({\bf r}, t) $ give correlators of observables (see
Eq.(\ref{72})).  Obviously, $ Z\{u\}$ represents the functional
Laplace transformation of the dependence $ Z\{\eta \} $,
$\delta$--function reflects the condition (\ref {3}), the
determinant is Jacobian of the integration variable change from
$\zeta $ to $ \eta $.

\subsection{ Fluctuation amplitude as a  component of nilpotent field}

Further development of the field scheme proceed
depending on
the type of
connection between stochastic variable $\zeta$ and order parameter
$\eta$. For thermodynamic system, where the thermostat state
does not depend on  $\eta$,
the determinant in Eq.(\ref{7}) assumes
constant value that can be
chosen as unity.
Then, by using integral representation
for $ \delta $--function

\begin {equation} \delta\{x({\bf
r},t)\}=\int\limits_{-{\rm i}\infty}^{{\rm i}\infty}\exp\left(
-\int\varphi x{\rm d}{\bf r} {\rm
d}t\right)D\varphi
\label{8}
\end {equation}
with the ghost field $\varphi({\bf r}, t)$
and averaging over distribution  (\ref
{5}), we have the functional (\ref {7}) in the standard form

\begin {equation}
Z\{\eta({\bf r},t)\}=\int\exp\left[-S\{\eta({\bf r},t), \varphi({\bf
r},t)\}\right] D\varphi,
\label{9}
\end {equation}
where the action $ S = \int
{\cal L} {\rm d} {\bf r} {\rm d} t $ is measured in units
$S_s=\gamma^2(T/D)^3$ with the Lagrangian given by

\begin {equation}
L(\eta,\varphi)=(\varphi\dot\eta-\varphi^2/2) + \varphi(\delta
V/\delta \eta ).\label{10}\end {equation}

In order to obtain a canonical form of the Lagrangian (\ref {10})
let us introduce  the nilpotent field

\begin {equation}
\phi_\varphi=\eta+\vartheta\varphi
\label {11}
\end {equation}
with Bose components $\eta$, $\varphi$,
and nilpotent coordinate $\vartheta $ obeys the relations

\begin {equation}
\vartheta^2=0,\qquad \int {\rm d}\vartheta=0,
\qquad \int\vartheta {\rm d}\vartheta=1.
\label{12}
\end {equation}
As is shown in  Appendix A,  the  first bracketed expression in
Lagrangian (\ref {10}) takes the form
of  kinetic energy in the
Dirac field scheme \cite {8}:

\def\theequation{\arabic{equation}{a}}\setcounter{equation}{12}
\begin {equation}
\kappa={1\over 2}\int \phi D \phi {\rm d}\vartheta.
\label {13a}
\end{equation}
\def\theequation{\arabic{equation}}\setcounter{equation}{13}
\vspace{-0.3cm}\noindent\\
Hereafter indexes are suppressed.
The Hermite operator $ D $ is defined by equality

\begin {equation}
D_\varphi=-{\partial \over \partial
\vartheta} +\left( 1 - 2\vartheta {\partial\over \partial \vartheta} \right)
{\partial\over \partial t}
\label{14}
\end {equation}
and enjoys the
property (\ref {A.6}). On the other hand, the algebraic properties
(\ref {12}) of  coordinate $ \vartheta $ allow to rewrite
the last term in Eq.(\ref {10}) in the standard form of potential
energy (see Appendix A)

\def\theequation{\arabic{equation}{b}} \setcounter{equation}{12}
\begin{equation}
\pi=\int V(\phi){\rm d}\vartheta.
\label{13b}
\end {equation}
\def\theequation{\arabic{equation}}\setcounter{equation}{14}
\vspace{-0.3cm}\noindent\\
The resulting Lagrangian  (\ref{10}) of the Euclidean
field theory is

\begin {equation}
L\equiv \kappa+\pi=\int \lambda \ {\rm d}\vartheta,
\quad\lambda(\phi)\equiv {1\over 2}\phi D\phi+V(\phi).
\label {15}
\end {equation}

According to Appendix A, the expressions (\ref{10}), (\ref{15})
become invariant with respect to  transformation
$e^{\varepsilon D}$  given by operator (\ref{14})  if only a
parameter $\varepsilon \to 0$ is pure imaginary and the fields
$\eta({\bf r},t)$, $\varphi({\bf r},t)$ are complex--valued. Then,
operator $D$ is the generator of the  nilpotent group.

After equating the first variation of the functional

\begin {equation}
s\{\phi(\zeta)\}=\int\lambda(\phi (\zeta)){\rm d}\zeta,
\qquad  \zeta\equiv \{{\bf r},t,\vartheta\}
\label {16} \end {equation}
to zero, we obtain the Euler--Lagrange equation


\begin {equation}
D{\delta \lambda\over \delta D
\phi}+{\delta \lambda\over \delta \phi}=0,
\label {17}\end{equation}
Substituting the expression (\ref{15}) in Eq.(\ref{17}) yields
the equation of motion

\begin{equation}
D\phi +\delta V/ \delta \phi=0.
\label{18}\end{equation}

Projection along axes of usual
and nilpotent   variables gives the system of the equations

\begin{equation}
\dot\eta=-\delta V/\delta \eta+\varphi,
\label {19}
\end {equation}
\begin {equation}
\dot\varphi={\delta^2 V\over \delta^2 \eta}\ \varphi,
\label {20}
\end{equation}
that determines kinetics of the  phase transition. Being
obtained from the extremum  condition for Lagrangian
(\ref{10}) these equations determine  the maximum value
of the probability  distribution

\begin {equation}
P\left\{\eta({\bf
r},t),\varphi ({\bf r},t)\right\}=Z^{-1}\exp\left(-\int
L(\eta,\varphi){\rm d} {\bf r}{\rm d} t\right) ,
\label {21}
\end{equation}
that specifies the partition function  $Z\equiv Z \{u = 0 \} $
in Eq.(\ref{6}). Comparison of expression (\ref {19}) with
Langevin equation (\ref {3}) leads to the conclusion that the
quantity $ \varphi $ determines the most probable value of fluctuation
of the field conjugated to the order parameter.
On the other hand, it means that the initial one--modal
distribution (\ref{5})  transforms
into the final two--modal form (\ref{21}).

\subsection{Conjugate field as a  component of nilpotent field}

It is well to bear in mind that there is another representation
of two--component nilpotent field. Let us introduce field
$f({\bf r},t)$ defined by the relation

\begin{equation}
\dot\eta=f+\varphi.
\label{22}
\end{equation}
Then the Lagrangian (\ref{10}) takes the form

\begin{equation}
L(\eta,f)={1\over
2}\left(\dot \eta^2 -f^2 \right) -{\delta V\over
\delta \eta}f  + {\delta V\over \delta \eta}\dot\eta.
\label{23}
\end{equation}
Since the last term of Eq.(\ref{23}) is the total derivative of
$V$ with respect to time, its contribution to the partition function
gives a factor that is integral over initial and final fields
$\eta_i ({\bf r})\equiv\eta({\bf r},t_i)$,
$\eta_f ({\bf r})\equiv\eta({\bf r},t_f)$
(here we return to dimensional magnitude of the potential
$V$).

\begin{equation}
Z=\int \exp \left(-
{V\{\eta_f\} - V\{\eta_i\} \over T} \right) D\eta_i D\eta_f
\label{24}
\end{equation}
The remaining part of Lagrangian (\ref{23}) yields the Euler
equations

\begin{equation}
\ddot\eta=-{\delta^2 V\over
\delta^2\eta}\ f,
\label{25}
\end{equation}
\begin{equation}
f=-\delta V/ \delta\eta.
\label{26}
\end{equation}
Differentiating Eq.(\ref{19}) with respect to time and taking into
account Eqs.(\ref{20}), (\ref{22}), it is not difficult to derive
Eq.(\ref{25}). As for Eq.(\ref{26}), it defines
$f({\bf r},t)$ as the field conjugated to the order
parameter $\eta({\bf r},t)$. Note that Eq.(\ref{26}) implies the force
$f$ explicitly does not depend on the time $t$.

By analogy with the definition (\ref{11}) let us introduce now
another nilpotent field \cite {9}

\begin{equation}
\phi_f=\eta - \vartheta f,
\label{27}
\end{equation}
where Bose components are the order parameter $\eta$ and the force
$f$ with opposite sign.  As it is shown in Appendix
A, expression for Lagrangian in terms of $\phi_f$ has the same form
as in Eq.(\ref{15})
with the generator of the nilpotent group given by

\begin{equation}
D_f=-\left( {\partial\over \partial \vartheta} +\vartheta
{\partial^2 \over \partial t^2}\right).
\label{28}
\end{equation}
Note that $D_f$ obeys the algebraic relation
(\ref{A.6}).

\subsection{ Connection between two--component nilpotent
representations }

In this subsection we discuss the relation between
the two above
two--component nilpotent fields
(\ref{11}) and (\ref{27})
that makes using of the fields algebraically equivalent.

Let us introduce the operators $\tau_\pm=e^{\pm
\vartheta \partial_t}$,  $\partial_t \equiv \partial/\partial t$
that induce the following transformations of  the fields
$\phi_\varphi$ and $\phi_{f}$

\begin{equation}
\tau_\pm\phi_{\mp\varphi}(t)=\phi_{\mp f}(t) ,\qquad
\tau_\pm\phi_{\pm f}(t)=\phi_{\pm \varphi}(t).\label{29}
\end{equation}
Eq.(\ref{29}) shows that operators $\tau_\pm$ transform
the field to its counterpart.
So we have the mappings relating the representations.

By making expansion in power
series over $\vartheta$, with help of Eqs.(\ref{12}), (\ref{22})
one obtaines

\begin{equation}
\phi_{\mp\varphi}(t\pm \vartheta)=\phi_{\mp f}(t) ,\qquad
\phi_{\pm f}(t\pm \vartheta)=\phi_{\pm \varphi}(t),\label{30}
\end{equation}
that shows that operators
$\tau_\pm$ shift the physical time $t$ by the nilpotent values
$\pm\vartheta$:

\begin{equation}
\tau_\pm\phi_{\mp\varphi}(t)=\phi_{\mp \varphi}(t\pm \vartheta) ,\qquad
\tau_\pm\phi_{\pm f}(t)=\phi_{\pm f}(t\pm \vartheta).\label{31}
\end{equation}
The same results can  be obtained by using
matrix representation defined by Eqs. (\ref{A.7}), (\ref{A.9})
and (\ref{A.10}).

On the other hand, the above mappings
$\tau_+\phi_f= \phi_\varphi$, $\tau_-\phi_\varphi=\phi_f$ induce
corresponding transformations of the generators
(\ref{14}), (\ref{28})

\begin{equation}
D_f=\tau_-D_\varphi \tau_+, \qquad D_\varphi=\tau_+D_f
\tau_-.\label{32} \end{equation}
Note that  the
action with   Lagrangian (\ref{15}) is
covariant with respect to
transformations (\ref{29}), provided
$\dot f\equiv 0$ (potential $V$ does not depend on time explicitly).
\section{Reduction of proper SUSY fields to the two--component
forms}

The considerations given in previous section rest on the
assumption that the Jacobian of variable change
from $\zeta$ to the order parameter $\eta$ is
constant. However, in general case determinant of an
arbitrary matrix $|A|$ can be expressed as an integral
over
Grassmann conjugate
fields $\psi({\bf r}, t)$, $\overline\psi ({\bf r}, t)$, that meet
conditions type of Eqs.(\ref{12})
 
\begin{equation}
\det |A|=\int  \exp\left( \overline\psi A\psi \right){\rm d}^2 \psi,
\qquad {\rm d}^2\psi = {\rm d} \psi {\rm d}\overline \psi.
 \label{33}
\end{equation}
Physically, the
appearance of new degrees of freedom $\psi$, $\overline\psi$ means that the
state of thermostat  turns out to be dependent on the order parameter ---
as it is inherent in self--organized system \cite {10}. As a result,
the Lagrangian (\ref{10}) supplemented with the Grassmann fields $\psi$,
$\overline\psi$ takes the form

\begin{equation}
{\cal L}(\eta,\varphi,\psi,\overline\psi)= \left(\varphi\dot\eta -
{\varphi^2\over 2} + {\delta V\over \delta \eta}\varphi\right) -
\overline\psi \left( {\partial \over \partial t}+{\delta^2 V \over
\delta \eta^2} \right)\psi.
 \label{34}
\end{equation}

Introducing the four--component SUSY field

\begin{equation}
\Phi_\varphi=\eta +\overline\theta\psi +\overline\psi\theta +
\overline\theta\theta \varphi,
 \label{35}
\end{equation}
by analogy with previous section the SUSY Lagrangian is

\begin{equation}
{\cal L} =\int \Lambda {\rm d}^2 \theta,\qquad \Lambda (\Phi_\varphi)
\equiv {1\over 2}(\overline {\cal D}_\varphi  \Phi_\varphi )\left(
{\cal D}_\varphi \Phi_\varphi\right) + V(\Phi_\varphi), \qquad {\rm
d}^2 \theta \equiv {\rm d} \theta{\rm d}\overline\theta,
 \label{36}
\end{equation}
where $\theta$, $\overline\theta$
are Grassmann conjugate coordinates that replace the nilpotent one
$\vartheta$.  As compared with Eq.(\ref{15}), where the kernel
$\lambda$ is linear in the generator (\ref{14}), a
couple of the Grassmann non--conjugated operators

\begin{equation} {\cal D}_\varphi={\partial \over
\partial \overline \theta}-  2\theta {\partial \over \partial t},
\qquad \overline {\cal D}_\varphi={\partial \over \partial \theta}
 \label{37}
\end{equation}
enters the expression for SUSY Lagrangian.
The  Euler  equation for SUSY action reads

\begin{equation}
-{1\over 2} [\overline {\cal D}, {\cal
D}]\Phi+{\delta V\over \delta \Phi}=0,
 \label{38}
\end{equation} where the square brackets denote
the commutator.  Projection of Eq.(\ref{38}) along the SUSY axes $1$,
$\overline\theta$, $\theta$, $\overline\theta\theta$ gives the
equations of motion

\def\theequation{\arabic{equation}{a}}\setcounter{equation}{38}
\begin{equation}
\dot\eta -\nabla^2\eta=-{{\partial} V/ {\partial}\eta} +
\varphi,
 \label{39a}
\end{equation}
\def\theequation{\arabic{equation}{b}}\setcounter{equation}{38}
\begin{equation}
\dot\varphi +\nabla^2\varphi = ({{\partial}^2 V/
{\partial}\eta^2}) \varphi -({{\partial}^3 V/
{\partial}\eta^3})\overline\psi\psi,
 \label{39b}
\end{equation}
\def\theequation{\arabic{equation}{c}}\setcounter{equation}{38}
\begin{equation}
\dot\psi -\nabla^2\psi =- ({{\partial}^2 V/
{\partial}\eta^2}) \psi,\label{39c}
\end{equation}
\def\theequation{\arabic{equation}{d}}\setcounter{equation}{38}
\begin{equation}
-\dot{\overline\psi}-\nabla^2\overline\psi=-({{\partial}^2 V/
{\partial}\eta^2}) \overline\psi,
\label{39d}
\end{equation}
\def\theequation{\arabic{equation}}\setcounter{equation}{39}
\vspace{-0.3cm}\noindent\\ that give Eqs.(\ref{19}),
(\ref{20}) at $\psi=\overline\psi=0$. It can be readily shown that
this system can be obtained from the Lagrangian (\ref{34}).
From Eqs.(\ref{39c}) and (\ref{39d})
we obtain the conservation law $\dot S + \nabla {\bf
j}=0$ for the quantities

\begin{equation}
S=\overline\psi\psi, \qquad
{\bf j}=(\nabla\overline\psi)\psi - \overline\psi(\nabla\psi).
 \label{40}
\end{equation}
For inhomogeneous thermodynamic
systems $S$ is a density of sharp boundaries, ${\bf j}$ is a
corresponding current \cite{6}. In  particular, the approach of the
four--component SUSY field complies with the strong  segregation limit
requirement of copolymer theory \cite{11}.  For self--organized system
the magnitude $S$ gives the entropy, ${\bf j}$ is the probability
current \cite{10}. So, for thermodynamic system, where the
entropy is conserved, we could  disregard the Grassmann fields
$\psi({\bf r},t)$, $\overline\psi ({\bf r}, t) = const$. As a result, the
four--component SUSY field (\ref{35}) is reduced to the
two--component form (\ref{11}).

In order to justify this let us write the kinetic term of the
Lagrangian (\ref{36}) in the form\break $-(1/4)
\Phi_\varphi\left[\overline {\cal D}_\varphi, {\cal D}_\varphi
\right]\Phi_\varphi$ where

\begin{equation}
-{1\over 2}\left[\overline{\cal D}_\varphi,{\cal D}_\varphi
\right]=-{\partial^2\over \partial\theta\partial\overline\theta}
+\left(1-2\theta{\partial \over \partial \theta} \right) {\partial
\over \partial t}.
 \label{41}
\end{equation}
The expression (\ref{41}) 
restricted to two--component form with
$\vartheta\equiv \overline\theta\theta$ 
yields
the generator (\ref{14}) as is needed. It is of interest to note that
variable $\vartheta$ satisfies  
(\ref{12}). 
In addition, since
the
self--conjugated value  $ \vartheta =
\overline\vartheta $
is commutating quantity, it  
is nilpotent rather than Grassmannian.

As in the case of  two--component nilpotent fields in Section II,
one can go over from the fluctuation amplitude $\varphi$ to the
conjugate force $f$ by using Eq.(\ref{22}). Then, the first bracket in
Lagrangian (\ref{34}) takes the form (\ref{23}) and instead of the
system (39) one obtains the equation (cf. Eq.(\ref{25}))

\begin{equation}
\ddot \eta = -(\delta^2 V/\delta \eta^2) f- (\delta^3
 V/\delta\eta^3)\overline\psi\psi
 \label{42}
\end{equation}
supplemented with the definition of force (\ref{26}) and the
equations (\ref{39c},d) for the Grassmann fields $\psi({\bf r}, t)$,
$\overline\psi({\bf r}, t)$. As above, the equation of motion (\ref{42})
can be derive by differentiating Eq.(\ref{39a}) with respect
to time and taking into account Eqs.(\ref{22}), (\ref{39b}). The
corresponding Lagrangian ${\cal L}(\eta,f,\psi,\overline \psi)$ takes
the SUSY form (cf. Eqs.(\ref{36}))

\begin{equation}
{\cal L} = \int \Lambda {\rm d}^2\theta, \qquad \Lambda(\Phi_f)\equiv
-{1\over 2} \Phi_f\overline{\cal D}_f{\cal D}_f \Phi_f + V(\Phi_f)
 \label{43}
\end{equation}
with the SUSY field  (cf. Eq.(\ref{27}))

\begin{eqnarray}
&&\Phi_f=\eta+\overline\theta\psi + \overline\psi\theta
-\overline\theta\theta f \equiv \Phi_\varphi
-\overline\theta\theta\dot\Phi_f=T_-\Phi_\varphi,
\nonumber \\
&&T_\pm \equiv e^{\pm \overline\theta\theta
\partial_t}, \quad \partial_t\equiv {\partial / \partial
t}\label{44} \end{eqnarray}
and the Grassmann conjugated operators (cf.  Eqs.(\ref{37}))

\begin{equation}
{\cal D}_f ={\partial \over \partial
\overline\theta}-\theta{\partial\over \partial t},\qquad
\overline{\cal D}_f={\partial \over \partial
\theta}-\overline\theta{\partial\over \partial t}.
 \label{45}
\end{equation}
By analogy with
Eqs.(\ref{29})--(\ref{31}) with operators
$\tau_\pm = e^{\pm \vartheta\partial_t}$
replaced by $T_\pm \equiv e^{\pm
\overline\theta\theta\partial_t}$, where $\partial_t\equiv
\partial/\partial t$,
the  SUSY fields (\ref{35}),
(\ref{44}) can be transformed into each other 
and  couples of the SUSY
operators (\ref{37}), (\ref{45}) are related by means of 
transformations (cf. Eqs.(\ref{32})):

\begin{equation}
{\cal D}_f=T_-{\cal D}_\varphi T_+, \qquad \overline {\cal D}_f=
T_-\overline{\cal D}_\varphi T_+.
 \label{46}
\end{equation}
According to Eq.(\ref{45}), kernel of kinetic part of 
the SUSY Lagrangian
(\ref{43}) is (cf. Eq.(\ref{41}))

\begin{equation}
-\overline{\cal D}_f {\cal D}_f = -\left( {\partial \over
\partial\theta}{\partial\over \partial \overline \theta}
+\overline \theta\theta {\partial^2\over \partial t^2}\right)+
\left( \overline \theta{\partial\over \partial\overline\theta} -
\theta {\partial\over \partial\theta} \right) {\partial\over
\partial t}.
 \label{47}
\end{equation}
Note that  the operator (\ref{28}) can be obtained 
from Eq.(\ref{47}) by
taking into account the condition
of the Fermion number conservation
$\overline\theta(\partial/\partial\overline\theta)=\theta(\partial/
\partial\theta)$ 
and by setting $\overline
\theta\theta$ equal to $\vartheta$.

So, both four--component Grassmann fields (\ref{35}), (\ref{44})
with  SUSY generators given by Eqs.(\ref{37}),
(\ref{45}) can be reduced to the
corresponding two--component fields,
Eqs.(\ref{11}), (\ref{27}), with operators (\ref{14}), (\ref{28}),
respectively.

It is worthwhile to mention that such reduction can be obtained
 according to the  SUSY gauge conditions

\begin{equation}
{\cal D}\Phi=0;\qquad \overline {\cal D}\Phi=0.
 \label{48}
\end{equation}
Indeed, according to  definitions (\ref{35}), (\ref{37}),
(\ref {44}), (\ref {45}) the equalities (\ref {48}) give the  relation

\begin{equation}
\overline\theta\psi+\overline\psi\theta -2\overline\theta\theta
f=0, \label{49}
\end{equation}
that reduces the  SUSY field  (\ref {44}) to the form (\ref{27}) with
 opposite sign before $f$, provided $\vartheta\equiv
\overline\theta\theta$.

Despite of
the same number of components,  one has to have in mind that 
the reduced SUSY field from 
Eq.(\ref{27}) 
and couple of Grassmann conjugate chiral SUSY fields (\ref
{B.9}), which appearance  is a consequence of SUSY gauge
invariance also (see Appendix B), have
different physical meaning.
The main distinction  is that the first
field consists of two Bose components $ \eta $, $ f $, whereas the
chiral SUSY fields $ \phi _ +$, $\phi_-$ are the combinations of
Bose $ \eta $ and Fermi $ \psi$,  $\overline\psi$ components.
Formally, this is due to the fact that  for separation of the
chiral SUSY fields the conditions (\ref {B.7}) of the  SUSY gauge
invariance are fulfilled not for the initial SUSY field $ \Phi $,
which satisfies to conditions (\ref {48}), but for components $ \Phi
_ \pm $, resulting from $ \Phi $ under the 
action of operators $ T _ \pm = \exp
\left(\pm \overline\theta\theta\partial_t \right) $ (see Eq.(\ref
{B.1})).

According to the above considerations, 
the transformation operators $T_\pm$,
that shift  physical time $t$ by Grassmann values $\pm\overline
\theta\theta$, relate the SUSY fields (\ref{35}), (\ref{44}) and
corresponding generators (\ref{37}), (\ref{45}). It should be
emphasized
that only the latter form a pair of  Grassmann conjugated operators. 
The physical
reason of this symmetry is that the corresponding equation of
(\ref{42})  is invariant with respect to the time inversion, whereas
the equations (\ref{39a}), (\ref{39b}) for components of the SUSY
field (\ref{35}) are not. However,in addition to the field
$\Phi_\varphi\equiv\Phi_+$  obtained from the initial field
$\Phi_f$ under the action of operator $T_+$, another SUSY field $\Phi_-$
emerge under the action of operator $T_-$ that 
shifts the time $t$ in opposite direction.
From Eqs.(\ref{B.5}), (\ref{22}) it can be seen that the fields
$\Phi_\pm\equiv\Phi_\varphi(\pm t)$ correspond to opposite
directions of time.

However,
equations (\ref{39c}), (\ref{39d}) for the Grassmann components
$\psi({\bf r}, t)$, $\overline\psi({\bf r},t)$ 
are invariant under the action of $T_\pm$ 
To break the invariance let us
introduce additional operators of  transformation

\begin{equation}
\widetilde T_\pm = \exp \left[ \varepsilon  \left(
\delta_\pm\overline\theta\psi + \delta_\mp\overline\psi\theta\right)
\right]
 \label{50}
\end{equation}
where source parameter $\varepsilon\to 0$;
$\delta_+=1$, $\delta_-=0$ for the positive time direction and
$\delta_+=0$, $\delta_-=1$  otherwise.
The Euler SUSY equation (\ref{38}) for transformed
superfield $\widetilde \Phi_\pm\equiv \widetilde T_\pm \Phi_\varphi$
is reduced to the components

\def\theequation{\arabic{equation}{a}}\setcounter{equation}{50}
\begin{equation}
\dot\eta -\nabla^2\eta=-\partial V/ \partial\eta +
\varphi-\varepsilon\overline\psi\psi,
 \label{51a}
\end{equation}
\def\theequation{\arabic{equation}{b}}\setcounter{equation}{50}
\begin{equation}
\dot\varphi +\nabla^2\varphi = (\partial^2 V/\partial\eta^2)
\varphi -(\partial^3 V/ \partial\eta^3)\overline\psi\psi +
\varepsilon\overline\psi\dot\psi,
\label{51b}
\end{equation}
\def\theequation{\arabic{equation}{c}}\setcounter{equation}{50}
\begin{equation}
\dot\psi -\nabla^2\psi =- (\partial^2 V/ \partial\eta^2)
\psi   -\varepsilon \left\{\delta_- (\dot\psi/\psi)\eta
+\delta_+\left[(\dot\eta-\varphi)+(\partial^2 V/
\partial\eta^2)\eta\right] \right\}\psi,
\label{51c}
\end{equation}
\def\theequation{\arabic{equation}{d}}\setcounter{equation}{50}
\begin{equation}
\dot{\overline\psi}+\nabla^2\overline\psi=({{\partial}^2 V/
{\partial}\eta^2}) \overline\psi-
\varepsilon \left\{ \delta_+(\dot{\overline\psi}/\overline\psi)\eta
-\delta_-\left[(\dot\eta-\varphi)+({{\partial}^2 V/
{\partial}\eta^2})\eta\right] \right\}\overline\psi,
 \label{51d}
\end{equation}
\def\theequation{\arabic{equation}}\setcounter{equation}{51}
\vspace{-0.3cm}\noindent\\ where the terms of first order 
 $\varepsilon$ are
kept.   These equations give
Eqs.(39)  at $\varepsilon\to 0$, 
but combination of  Eqs.(\ref{51c}), (\ref{51d}) at
$\varepsilon\ne 0$ leads to the following equation  for 
the quantities (\ref{40})

\begin{equation}
\dot S+\nabla{\bf j}=\pm \varepsilon F S,\quad F\equiv \partial
V/ \partial\eta -2(\partial^2 V/ \partial\eta^2)\eta
\label{52}
\end{equation}
instead of the law of entropy conservation.
Since  entropy $S$ of a closed system ($\nabla {\bf j}=0$) 
increases in time, provided
$F >0$, in Eq.(\ref{52}) one has to choose
the upper sign corresponding  to the positive time direction. So, the
operator (\ref{50}) breaks symmetry with respect to the time
reversibility. The above--mentioned condition of positiveness  for
effective force  $F\equiv \partial V/ \partial\eta
-2(\partial^2 V/ \partial\eta^2)\eta $ means that the  effective
potential $V$ is an increasing convex function of the
$\eta$ 
that is inherent in an unstable system.
It is of interest to note that   near
the  equilibrium state, where $\partial V/ \partial \eta=0 $,
$\eta\ll 1$, the force $F\simeq - (\partial^2 V/ \partial\eta^2)\eta$
is always positive for unstable system.

Finally, in order to  visualize the difference between
two--component nilpotent fields (\ref {11}),
(\ref{27}) and chiral fields (\ref {B.9}) 
let us represent the SUSY
field (\ref {44}) as a vector in four--dimensional  space with
axes  $ \theta ^ 0 = \overline\theta ^ 0\equiv 1 $, $ \overline\theta
$, $ \theta $, $ \overline\theta\theta\equiv \vartheta $.  Then
conditions (\ref{48}) of the SUSY gauge invariance mean that
field (\ref {44}) is reduced to the vector (\ref {27})
belonging to a plane formed by axes 1, $ \vartheta $.  Accordingly,
the conditions (\ref {B.7}) of the chiral  gauge invariance 
split total SUSY space into a couple of orthogonal subspaces, the
first of which has the  axes 1, $ \theta $  and contains the
vector $ \phi _ - $, and second --- axes 1, $ \overline\theta $ and
vector $ \phi _ + $. Since these subspaces are
Grassmann conjugated, $ \overline\phi _ - = \phi _ + $, it is enough
to use one of them, considering either vector $ \phi _ - $, or $ \phi
_ + $ (see Appendix B).  Such program was realized in  
Ref.\cite{12}, whereas the above used nilpotent field 
(\ref {27}) is derived by
projecting  chiral vectors $ \phi _ \pm $ to a plane formed by axes
1, $ \vartheta $.  It follows that our approach stated on the
using nilpotent fields (\ref{11}), (\ref{27}) and the theory \cite {12}
are equivalent.  The SUSY method presented in the book \cite{13}  
is also based
on usage  of the chiral fields
$ \phi _ - = \varphi -i\overline\psi\theta $, $ \phi _ +
 = \eta + \overline\theta\psi $ (cf. with (\ref {B.9})) that contain
the fluctuation
$\varphi$ as a Bose component of the  field $ \phi _ - $ and
the order parameter  $ \eta $ in  field $\phi_+$.

\section{SUSY correlation techniques}

In this section  correlators
of the proper SUSY fields (\ref{35}), (\ref{44}) will be studied.
It will be shown how the relevant correlation
techniques can be reduced to the simplest scheme by
making use of the two--component
field (\ref{11}).

To begin with let us introduce the SUSY correlator

\begin{equation}
C(z,z')= \langle\Phi(z)\Phi(z')\rangle,\qquad
z\equiv\{{\rm{\bf r}},t,\overline\theta,\theta\}.
\label{53}\end{equation}
From the equation of motion (\ref{38}) we have
the equation for Fourier transform of
the bare SUSY correlator
$C^{(0)}(z,z')$ with the potential $V_0=(1/2)\Phi^2$
in the following form

\begin{equation}
L_{\rm{\bf k}\omega}(\theta)C_{{\bf
k}\omega}^{(0)}(\theta,\theta ')=\delta(\theta,\theta '),\qquad
L\equiv 1-(1/ 2) [\overline {\cal D},{\cal D}],\label{54}
\end{equation}
where
$\delta(\theta,\theta')$ is the
Grassmann $\delta$--function

\begin{equation}
\delta(\theta,\theta')=(\overline\theta-\overline\theta')
(\theta-\theta'),\label{55} \end{equation}

$\omega$ is the frequency
and ${\bf k}$ is the  wave vector.
The
solution of Eq.(\ref{54}) reads

\begin{equation}
C^{(0)}(\theta,\theta ')={\left(1+(1/2)[\overline
{\cal D},{\cal D}]\right)\delta(\theta,\theta ')\over
1-(1/4)[\overline {\cal D}, {\cal D}]^2},
\label{56}\end{equation}
where the indexes $\omega$, ${\bf
k}$ are suppressed for brevity.  From the definitions
(\ref{37}), (\ref{55}) and equality $[\overline {\cal
D}, {\cal D}]^2=-4\omega^2$ (see Eqs.(\ref{A.11})),
the bare SUSY correlator for
SUSY field (\ref{35}) can be written in
the explicit form

\begin{equation}
C_\varphi^{(0)}(\theta,\theta
')={1+ (1-{\rm
i}\omega)(\overline\theta - \overline\theta') \theta  - (1+{\rm
i}\omega)(\overline\theta - \overline\theta')  \theta'\over
1+\omega^2}. \label{57}\end{equation}
In the case of the SUSY field (\ref{44}),
by using transformation (\ref{46}) the above result is found
to be modified by
adding the term
${\rm i}\omega (\overline\theta\theta -\overline\theta'\theta')$ to
the numerator of Eq.(\ref{57}).

It is convenient to introduce the following components
as a basis for expansion of SUSY correlators

\begin{eqnarray}
&&T(\theta,\theta ')=1,\qquad B_0(\theta,\theta
')=\overline\theta\theta,\qquad B_1(\theta,\theta ')=\overline\theta
   '\theta ',\label{58}\\
&&F_0(\theta,\theta
   ')=-\overline\theta '\theta,\qquad F_1(\theta,\theta
')=-\overline\theta\theta '. \nonumber \end{eqnarray}
Let us define the  operator product

\begin{equation}
X(\theta,\theta ')=\int Y(\theta,\theta '')Z(\theta '',\theta ')
{\rm d}^2 \theta '' \label{59}
\end{equation}
for superspace functions $ Y$, $ Z$.
Eq.(\ref{59}) immediately provide  the
multiplication rules for the basis operators (\ref{58})
summarized in Table I:

\hspace{7.3cm} {Table I}
\begin{center}
\begin{tabular}[h]{|c||c|c|c||c|c|c|}
\hline &&&&&\\[-18pt]
{$l\backslash r$} &${\bf T}$ & ${\bf B}_0$ & ${\bf B}_1$
&${\bf F}_0$ & ${\bf F}_1$ \\
\hline &&&&& \\[-20pt]
\hline &&&&& \\[-18pt]
${\bf T}$ & 0 &${\bf T}$ & 0 &   0 &  0 \\
\hline &&&&& \\[-18pt]
${\bf B}_0$ & 0 &${\bf B}_0$ & 0  & 0 & 0 \\
\hline &&&&& \\[-18pt]
${\bf B}_1$ & ${\bf T}$ & 0 & ${\bf B}_1$ & 0 & 0 \\
\hline &&&&& \\[-42pt]
&&&&&\\
\hline &&&&& \\[-18pt]
${\bf F}_0$ & 0 &  0 & 0 &   ${\bf F}_0$ & 0 \\
\hline &&&&& \\[-18pt]
${\bf F}_1$ & 0 &  0 &  0 & 0 & ${\bf F}_1$\\ \hline \end{tabular}
\end{center} 

\vspace{-0.3cm}\noindent\\ The operators ${\bf T}$, ${\bf B}_{0,1}$,
${\bf F}_{0,1}$  then  form the closed basis,
so that
expansions for  correlators are
(see Eqs.(\ref{C.7}), (\ref{C.10}))

\begin{eqnarray}
&&{\bf C}_\varphi=S{\bf T}+ G_+\left({\bf B}_0 +{\bf F}_0\right)
   +G_-\left({\bf B}_1+{\bf F}_1 \right),\label{60}\\
&&{\bf C}_f=S{\bf T}+ m_+{\bf B}_0+m_-{\bf B}_1+G_+{\bf
F}_0+G_-{\bf F}_1 \nonumber\end{eqnarray}
where in  accordance with Ward identity (\ref{C.6}) corresponding
to the first generator (\ref{C.5})  term  proportional to
$\overline\theta\theta\overline\theta'\theta'$ is dropped.
Inserting SUSY fields (\ref{35}), (\ref{44})   into
Eq.(\ref{53}) provides the coefficients of  expansions (\ref{60})
(cf. Eqs.(\ref{C.8}), (\ref{C.11})):

\begin{eqnarray}
&&S=\langle |\eta|^2\rangle; \quad  m_+=\langle \eta^*\rangle
f_{\rm ext},\quad m_-=\langle \eta \rangle f^*_{\rm ext},\quad
f_{\rm ext}\equiv -f; \label{61}\\
&&G_+=\langle
\varphi\eta^*\rangle= \langle \overline\psi\psi^*\rangle,\quad
G_-=\langle \eta\varphi^*\rangle=\langle
\overline\psi^*\psi\rangle.\nonumber
\end{eqnarray}
So, quantity $S$ is the autocorrelator of order parameter $\eta$ and
magnitudes $m_\mp$ meet the condition $m_+^*=m_-$ and  determine
the averaged order parameter $\left<\eta\right>$  corresponding to
external force $f_{\rm ext }\equiv -f$. The retarded and advanced
Green functions $G_\mp$ give the response of order parameter $\eta$
to fluctuation amplitude $\varphi$  and {\it vice versa} (moreover,
functions $G_\pm$ determine correlation of the Grassmann fields
$\overline\psi$, $\psi$).  As it is known \cite{14}, the  Fourier
transforms $G_\mp (\omega)$ of retarded and advanced Green functions
are analytical in upper and lower half--planes of complex frequency
$\omega$  with cut along real axis $\omega'$. There is the
jump $G_-(\omega') - G_+(\omega')=4{\rm i}\, {\rm  Im}\,
G_-(\omega')$,  so that the relations (\ref{C.9}), (\ref{C.12})
assume the usual form of the fluctuation--dissipation theorem:

\begin{equation}
G_\pm(\omega) =m_\pm (\omega) \mp {\rm i}
\omega S(\omega),\quad
S(\omega') =(2/\omega')\, {\rm Im}\, G_-(\omega')
\label{62} \end{equation} where the
frequency $\omega'$ is  real.  The
expression for bare correlator (\ref{57}) gives:

\begin{eqnarray}
&&S^{(0)}= m_{\pm}^{(0)}=(1+\omega^2)^{-1},\qquad
 G_\pm^{(0)}=(1\pm {\rm i}\omega)^{-1}.\label{63}
\end{eqnarray}

Integrate the last equation of
(\ref{62}) and taking into account the spectral representation

\begin{equation}
C(\omega)=\int_{-\infty}^\infty {{\rm d}\omega'\over \pi}{{\rm
Im}\, C(\omega')\over \omega'-\omega}. \label{64}
\end{equation}
we arrive at useful relation

\begin{equation}
S(t=0)=G_\pm(\omega=0)\equiv \chi, \label{65}
\end{equation}
where the last identity is the definition of susceptibility
$\chi$.

The expansions (\ref{60}) make it possible to handle the SUSY
correlator (\ref{53}) as a vector of space constructed as the direct
product of the SUSY fields (\ref{35}) or (\ref{44}).
The representation (\ref{35})
is of special convenience  because it
allows using of reduced basis

\begin{equation}
{\bf A}\equiv{\bf B}_0+{\bf F}_0,\qquad {\bf B}\equiv{\bf
B}_1+{\bf F}_1.\label{66} \end{equation}
Along with  ${\bf T}$, they form
more compact basis and  obey the following
multiplication rules:

\hspace{7.2cm} {Table II}
\begin{center}
\begin{tabular}[h]{|c||c|c|c|}
\hline &&&\\[-18pt]
{$l\backslash r$} & ${\bf T}$ & ${\bf A}$ & ${\bf B}$ \\
\hline &&& \\[-20pt]
\hline &&& \\[-18pt]
${\bf T}$ & $0$ & ${\bf T}$ & $0$\\
\hline &&& \\[-18pt]
${\bf A}$ & $0$ & ${\bf A}$ & $0$\\
\hline &&& \\[-18pt]
${\bf B}$ & ${\bf T}$ & $0$ & ${\bf B}$ \\
\hline
\end{tabular}
\end{center}

\vspace{-0.3cm}\noindent\\ The expansion   Eqs.(\ref{60}) then
takes the form

\begin{equation} {\bf C}_\varphi=S{\bf T} + G_+{\bf A}+G_-{\bf B}.
\label{67}\end{equation}

So, using Ward identities allows to get rid of
autocorrelators of the Grassmann fields $\psi$, $\overline\psi$
(see relations (\ref{61})).  As a result, there are three basic
correlators: the advanced  and retarded
Green functions  $G_\pm$
and structure factor $S$. They yield the
most compact expansion (\ref{67}) for arbitrary SUSY correlator of
fields (\ref{35}).
It is ready to show that  expansion of the same form can be
obtained on the basis of the two--component field (\ref{11})
representation.
Indeed, in this case
by comparison between equations of motion (\ref{18}) and (\ref{38})
the  commutator $-(1/2)[\overline {\cal D}, {\cal D}]$ in expression
(\ref{56}) should be replaced by generator (\ref{14}) and
nilpotent $\delta$--function should be
$\delta(\vartheta-\vartheta')=\vartheta+\vartheta'$.
So the resulting bare correlator is

\begin{equation}
C^{(0)}(\vartheta,\vartheta')={1+(1-{\rm i}\omega)\vartheta+(1+{\rm
i}\omega)\vartheta' \over 1+\omega^2}
\label{68}\end{equation}
instead of Eq.(\ref{57}). It is easily to see that using the
definitions (cf.  Eqs.(\ref{58}))

\begin{equation}
T(\vartheta,\vartheta')=1,\quad A(\vartheta,\vartheta')=\vartheta,
 \quad B(\vartheta,\vartheta')=\vartheta'
 \label{69}\end{equation}
gives the relevant expansion (\ref{67}). As a result, in what follows
we can use two--component field (\ref{11}).

In particular, for inverse of the SUSY correlator (\ref{67})
we have

\begin{equation} {\bf C}^{-1}= -G_+^{-1}
S G_-^{-1} {\bf T}+G_+^{-1} {\bf A} +G_-^{-1} {\bf B}.
\label{70} \end{equation}

It is worthwhile to note that according to definitions
(\ref{66}), (\ref{58}) the basis operators ${\bf A}\equiv {\bf
B}\equiv 0$ provided   $\theta=\theta'$, so that
$C(\theta,\theta)=C(\vartheta,\vartheta)= S$ and

\begin{eqnarray}
&&\int C(z,z){\rm d}z = \int S({\bf r}, t;{\bf r}, t)
{\rm d}{\bf r} {\rm d} t {\rm d}^2\theta=0,\label
{71}\\ &&\int C(\zeta,\zeta){\rm d}\zeta = \int S( {\bf r}, t;
{\bf r}, t){\rm d}{\bf r} {\rm d} t {\rm
d}\vartheta=0\nonumber \end{eqnarray}
where
 $z$, $\zeta$ are sets of variables
$\{ {\bf r}, t, \overline\theta, \theta
\}$, $\{ {\bf r}, t, \vartheta \}$, respectively.
In the diagrammatic representation
identities
(\ref{71}) imply the absence of bubble graphs  contribution.
  The latter considerably reduces
 the number of graphs contributing to
expansion of the perturbation theory (see below).

\section{SUSY Perturbation Theory}

Let us begin with the formula

\begin {equation}
C(\zeta,\zeta')=\left.  {\delta^2 Z\{ u(\zeta)\}\over
\delta u(\zeta)\delta u(\zeta')} \right|_{u=0},
\label{72}\end {equation} where generating functional (see Eqs.(\ref
{6}), (\ref {9}))

\begin{equation}
Z\{u\}=\left<\exp\left( \int \phi u {\rm d}\zeta
\right)\right>\label{73}\end{equation}
has the form of average over distribution (cf. Eq.(\ref {21}))

\begin{equation}
P\{\phi\}=Z^{-1}
\exp\left(-S\{\phi\}\right),\qquad
S\{\phi\}=\int \lambda (\phi){\rm
d}\zeta,\label{74}\end{equation}
with the Lagrangian $ \lambda
$ defined by Eq.(\ref {15}).  In the zero--order approximation
the action is quadratic

\begin{equation}
S_0={1\over 2}\int \phi L\phi {\rm d}\zeta, \qquad
L\equiv 1+ D,\label{75}\end{equation}
where generator $D$ is
given by Eq.(\ref{14}). Corresponding distribution takes the SUSY
Gaussian form (cf. Eq.(\ref{5}))

\begin{equation}
P_0\{\phi\}=\left({ \det|L| \over
2\pi}\right)^{1/2}\exp\left\{-{1\over 2}\int \phi L\phi {\rm
d}\zeta\right\}.
\label{76}\end{equation}
So for the bare
supercorrelator we have the expression

\begin{equation}
C^{(0)}(\zeta,\zeta')=L^{-1}\delta (\zeta,\zeta'),
\quad \delta(\vartheta,\vartheta') \equiv \vartheta+\vartheta',
\label{77}\end{equation}
that leads to Eq.(\ref{56}) with $-(1/2)[\overline{\cal D},
{\cal D}]$ replaced by $D$ if Eqs.(\ref{75}), (\ref{14})
are taken into
account.  The linear
operator ${\bf L}\equiv ({\bf C}^{(0)})^{-1}$ in accordance with
Eq.(\ref{70}) takes the form:

\begin{eqnarray} &&{\bf L}=L{\bf T} + L_+{\bf A}+L_-{\bf B};
\label{78}\\
&& L=-1,\quad  L_\pm=1 \pm i\omega. \nonumber
\end{eqnarray}

To proceed, one need to separate out anharmonic
part $S_1\{\phi\}$ of exponent in distribution (\ref{74})
as a perturbation
and
to make expansion in power series over $S_1$.
Insertion of this series into Eq.(\ref{72}) gives

\begin{eqnarray}
&&{\bf C}(\zeta,\zeta')=\sum_{n=0}^\infty
{(-1)^n\over n!} \left< \phi(\zeta)\left( S_1 \{ \phi\}
\right)^n \phi(\zeta')\right>_0, \label{79}\end{eqnarray}
where subscript "0" means  averaging  over the bare
distribution (\ref{76}). Further one has to make factorization
by making use of the Wick theorem. Then within the $n$--th order of
perturbation theory the expression (\ref{79}) takes the form

\begin {equation}
C^{(n)}(\zeta,\zeta ') = \int\!\!\int C^{(0)}(\zeta, \zeta_1)\Sigma^{(n)}
(\zeta_1,\zeta_2) C^{(0)} (\zeta_2,\zeta') {\rm d} \zeta_1\ {\rm d} \zeta_2,
\label{80}\end {equation}
where $\Sigma ^{(n)} (\zeta_1,\zeta_2)$ is the SUSY self--energy function of
$n$--th order that should be calculated. The result
essentially  depends on the form of
$V_1(\phi)$ that describes self--action effects. In what follows we will
analyse two widely used models.

\subsection {$\phi^4$--model}

Let the self--action potential be defined by the quartic
dependence

\begin{equation}
V_1(\zeta)={\lambda \over 4!}\phi^4(\zeta),\qquad \zeta=\{ {\bf r},
t,\vartheta\}\label{81}\end{equation}
with the anharmonicity constant
$\lambda >0$. Then terms of the first and second orders of series
(\ref{79}) are

\def\theequation{\arabic{equation}{a}}\setcounter{equation}{81}
\begin{equation}
C^{(1)}(\zeta,\zeta')=-{\lambda\over 4!}\int
\left<\phi(\zeta)(\phi(\zeta_1))^4\phi(\zeta')\right>_0
{\rm d}\zeta_1,\label{82a}
\end{equation}
\def\theequation{\arabic{equation}{b}}\setcounter{equation}{81}
\begin{equation}
C^{(2)}(\zeta,\zeta')={1\over 2!}
\left(-{\lambda \over  4!}\right)^2\int\!\!\int \left<\phi(\zeta)
(\phi(\zeta_1))^4 (\phi(\zeta_2))^4
\phi(\zeta')\right>_0 {\rm d}\zeta_1\ {\rm d}\zeta_2\ .
\label{82b}\end{equation}
\def\theequation{\arabic{equation}}\setcounter{equation}{82}
\vspace{-0.3cm}\noindent\\ Now one has to count the number of
possible pairings when using the Wick theorem. In
Eq.(\ref{82a})  the total number  of  pairings  is
12, and the formula (\ref {82a}) reads

\begin{equation}
C^{(1)}(\zeta,\zeta')=-{\lambda\over 2} \int C^{(0)}(\zeta,\zeta_1)
C^{(0)}(\zeta_1,\zeta_1) C^{(0)}(\zeta_1,\zeta'){\rm d}\zeta_1\equiv
0, \label{83}\end{equation}
where Eqs.(\ref{53}), (\ref{71}) are
taken into account.  In Eq.(\ref {82b}) the total number  of
pairings equals 192 and  the Wick theorem gives

\begin{equation}
C^{(2)}(\zeta,\zeta')={\lambda^2 \over 6}\int\!\!\int
C^{(0)}(\zeta,\zeta_1) \left(C^{(0)}(\zeta_1,\zeta_2)\right)^3
C^{(0)}(\zeta_2,\zeta'){\rm d}\zeta_1\ {\rm d}\zeta_2.
\label{84}
 \end{equation} Then, in accord with
Eq.(\ref{80}) the SUSY self--energy function in the second
order of perturbation theory reads

\begin{equation}
\Sigma(\zeta,\zeta')={\lambda^2\over 6}
\left(C(\zeta,\zeta')\right)^3.
\label{85}  \end{equation}
Here  in terms
of usual diagram ideology bare correlator is
replaced by exact one.

In the diagrammatic representation
terms (82a,b) correspond to the following graphs:

\begin{figure}[h]
\hspace{5cm}
\special{em:graph diag1.pcx}
\vspace*{2cm}
\end{figure}

\noindent According to the rule
(\ref{71}) the former does not contribute to correlator, whereas the
latter does (\ref{85}).

By analogy with the
SUSY correlator (\ref{67}) it is convenient to expand the SUSY
self--energy:

\begin{equation}
{\bf\Sigma}=\Sigma{\bf T}+\Sigma_+{\bf A}+\Sigma_-{\bf B}.
\label{86}
 \end{equation}
To determine the coefficients $\Sigma_\pm$, $\Sigma$ it should be
taken into account that the multiplication rules in Eq.(\ref{85})
differ from  ones given by Table II. The reason is that
Eq.(\ref{85}) contains  "element-to-element"  products of
nilpotent quantities \cite{3} instead of the above operator product.
Hence one has to use the
multiplication rules given by the Table III:

\hspace{7.2cm} {Table III}
\begin{center}
\begin{tabular}[h]{|c||c|c|c|}
\hline &&&\\[-18pt]
{$l\backslash r$}&$T (\vartheta,\vartheta')$ & $A(\vartheta,\vartheta')$ &
$B(\vartheta,\vartheta')$ \\
\hline &&& \\[-20pt]
\hline &&& \\[-18pt]
$T(\vartheta,\vartheta')$ & $T(\vartheta,\vartheta')$ &
$A(\vartheta,\vartheta')$ & $B(\vartheta,\vartheta')$\\
\hline &&& \\[-18pt]
$A(\vartheta,\vartheta')$ & $ A(\vartheta,\vartheta')$ & $0$ & $0$\\
\hline &&& \\[-18pt]
$B(\vartheta,\vartheta')$ & $B(\vartheta,\vartheta') $ & $0$ & $0 $\\
\hline \end{tabular} \end{center}

As a result, the coefficients of expansion (\ref{86}) take
the form:

\def\theequation{\arabic{equation}{a}}\setcounter{equation}{86}
\begin{equation}
\Sigma(t)=(\lambda^2/6) S^3(t), \label{87a}
\end{equation}
\def\theequation{\arabic{equation}{b}}\setcounter{equation}{86}
\begin{equation}
\Sigma_\pm(t)=(\lambda^2/2)S^2(t)G_\pm(t).\label{87b}
\end{equation}
\def\theequation{\arabic{equation}}\setcounter{equation}{87}
\vspace{-0.3cm}\noindent\\ In the frequency representation that will be
needed below we have

\def\theequation{\arabic{equation}{a}}\setcounter{equation}{87}
\begin{equation}
\Sigma(\omega)={\lambda^2\over 6}\int {{\rm d}\omega_1\ {\rm
d}\omega_2\over (2\pi)^2} S(\omega-\omega_1-\omega_2)
S(\omega_1)S(\omega_2),  \label{88a}
\end{equation}
\def\theequation{\arabic{equation}{b}}\setcounter{equation}{87}
\begin{equation}
\Sigma_\pm(\omega)={\lambda^2\over 2}\int {{\rm d}\omega_1\ {\rm
d}\omega_2\over (2\pi)^2} G_\pm(\omega-\omega_1-\omega_2)
S(\omega_1)S(\omega_2). \label{88b}
\end{equation}
\def\theequation{\arabic{equation}}\setcounter{equation}{88}
\vspace{-0.3cm}\noindent\\ The obvious inconvenience of this
expressions is the presence of convolutions. To get rid of them let us
use the fluctuation--dissipation  theorem

\begin{equation}
\Sigma(t=0)=\Sigma_\pm(\omega=0) \label{89}
\end{equation}
in the form of Eq.(\ref{65}).
Then from Eqs.(\ref{87a}), (\ref{65}) one obtains:

\def\theequation{\arabic{equation}{a}}\setcounter{equation}{89}
\begin{equation}
\Sigma_\pm(\omega=0)=(\lambda^2/6)\chi^3.
\label{90a}
\end{equation}
\def\theequation{\arabic{equation}}\setcounter{equation}{90}

\subsection{Cubic anharmonicity}

Apart from the $\phi^4$--model
studied above, there is a number of physical systems
type of copolymers \cite {2} where
cubic anharmonicity

\begin{equation}
V_1(\zeta)={\mu\over 3!} \phi^3(\zeta), \qquad \zeta=\{ {\bf r},
t,\vartheta\} \label{91}\end{equation}
 has a dominant role
( $\mu$ is the anharmonicity parameter).  By
analogy with Eqs.(82) it can be shown that the first
non--vanishing contribution to the SUSY
correlator (\ref {80}) is

\begin{equation}
 C^{(2)} (\zeta',\zeta')={1\over 2!}\left(-{\mu\over
3!}\right)^2\int\!\!\int
\left<\phi(\zeta)(\phi(\zeta_1))^3 (\phi(\zeta_2))^3 \phi(\zeta')
\right>_0 {\rm d}\zeta_1\ {\rm d}\zeta_2.
\label{92} \end{equation}
To facilitate the factorization
of these products let us depict  possible graphs of the second
order in cubic anharmonicity $\mu$:

\hspace{5cm}
\special{em:graph dig2.pcx}
\vspace*{2cm}

\noindent The first of these graphs contains the
bubble and does not contribute to the correlator.
The contribution of the second graph is

\begin{equation}
{\mu^2\over 2}\int\!\!\int C^{(0)} (\zeta,\zeta_1)\left(
C^{(0)}(\zeta_1,\zeta_2)\right)^2 C^{(0)} (\zeta_2,\zeta') {\rm d}\zeta_1\ {\rm d}\zeta_2.
\label{93} \end{equation}
As a result,  the SUSY
self--energy function reads

\begin{equation}
\Sigma(\zeta,\zeta')={\mu^2\over 2}
\left(C(\zeta,\zeta')\right)^2, \label{94} \end{equation}
where  the bare SUSY correlators are replaced by exact ones.
By using the multiplication rules from Table III
the coefficients
of the expansion (\ref{86}) are derived

\def\theequation{\arabic{equation}{a}}\setcounter{equation}{94}
\begin{equation}
\Sigma(t)= (\mu^2/2)S^2(t),\label{95a}
\end{equation}
\def\theequation{\arabic{equation}{b}}\setcounter{equation}{94}
\begin{equation}
\Sigma_\pm(t)=\mu^2 S(t)G_\pm(t).\label{95b}
\end{equation}
\def\theequation{\arabic{equation}}\setcounter{equation}{95}
\vspace{-0.3cm}\noindent\\
These expressions, combined with Eqs.(87),
determine the SUSY self-energy
function  completely. By analogy with Eq.(\ref{90a})
we have the relation

\def\theequation{\arabic{equation}{b}}\setcounter{equation}{89}
\begin{equation}
\Sigma_\pm(\omega=0)=\Sigma (t=0)\equiv (\mu^2/2)\chi^2
\label{90b} \end{equation}
\def\theequation{\arabic{equation}}\setcounter{equation}{89}

Finally,
the resulting expressions
for coefficients of expansion (\ref{86}) with
both cubic and quartic anharmonicities included are

\def\theequation{\arabic{equation}{a}}\setcounter{equation}{95}
\begin{equation}
\Sigma(t)={1\over 2}
\left(\mu^2 +  {\lambda^2 \over
3}S(t)\right)S^2(t),\label{96a}
\end{equation}
\def\theequation{\arabic{equation}{b}}\setcounter{equation}{95}
\begin{equation}
\Sigma_\pm(t)=\left(\mu^2 + {\lambda^2\over 2} S(t)\right)
S(t)G_\pm(t),\label{96b}
\end{equation}
\def\theequation{\arabic{equation}{c}}\setcounter{equation}{95}
\begin{equation}
\Sigma_\pm(\omega=0) = {1\over 2} \left(\mu^2 +  {\lambda^2 \over
3}\chi\right)\chi^2\label{96c}
\end{equation}
\def\theequation{\arabic{equation}}\setcounter{equation}{96}
\section{Self--consistent approach}

\subsection{Effective SUSY Lagrangian}

Let us start with the total SUSY action taken in the site
representation:

\begin{equation}
S=S_0+S_1+S_{int}; \label{97}
\end{equation}
\def\theequation{\arabic{equation}{a}}\setcounter{equation}{96}
\begin{equation}
S_0\equiv {1\over 2}\sum_{ l}\int\
\phi_{l}(t,\vartheta)
\Bigl[1+D(\vartheta)\Bigr] \phi_{l}(t,\vartheta){\rm d}t{\rm
d}\vartheta, \label{97a}
\end{equation}
\def\theequation{\arabic{equation}{b}}\setcounter{equation}{96}
\begin{equation}
S_1\equiv \sum_{l}\int\
V_1\left(\phi_{ l}(t,\vartheta)\right){\rm d}t{\rm d}\vartheta,
\label{97b}
\end{equation}
\def\theequation{\arabic{equation}{c}}\setcounter{equation}{96}
\begin{equation}
S_{int}\equiv \int \int\
V_{int}\left\{\phi_{
l}(t,\vartheta),\phi_{m}(t',\vartheta')
\right\}\delta(t-t')
{\rm d}t{\rm d}t'{\rm d}\vartheta{\rm d}\vartheta',
\qquad V_{int}\equiv V + W.  \label{97c}
\end{equation}
\def\theequation{\arabic{equation}}\setcounter{equation}{97}
\vspace{-0.3cm}\noindent\\ where sites are labeled with $l$ and
the self--action term $V_1(\phi_{l})$ (\ref{97b}),
that given by Eqs.(\ref{81}) and (91),
is separated out.
The last term
$S_{int}$ describes the two--particle interaction $V$ and
the effective potential $W$ is caused by averaging  over
quenched disorder.
The potential $V$ is assumed to be attractive and
takes the standard form \cite{15}

\begin{eqnarray}
&&V = - {1\over 2}\sum_{ lm} v_{lm} \phi_{m}(t,\vartheta)
\phi_{l}(t',\vartheta') \phi_{l}(t',\vartheta') \phi_{m}
(t,\vartheta)-\nonumber\\
&&\qquad {1\over 2} \sum_{lm} v_{lm} \phi_{l}(t,\vartheta)
 \phi_{l}(t',\vartheta') \phi_{m}(t,\vartheta)
\phi_{m}(t',\vartheta')
\label{98} \end{eqnarray}
that, in the mean--field approximation,
provides the following expression

\begin{eqnarray}
&&V\simeq -{v\over 2}  C(t,\vartheta;t,\vartheta)\sum_{l}
\phi_{l}(t',\vartheta') \phi_{l}(t',\vartheta')- \nonumber\\
&&\qquad {v\over 2}
C(t,\vartheta;t',\vartheta')\sum_{l}  \phi_{l}(t,\vartheta)
\phi_{l}(t',\vartheta'). \label{99}\end{eqnarray}
Hereafter $v\equiv \sum_{m} v_{lm} >0$  is the interaction  constant,
$ C(t,\vartheta;t',\vartheta')\equiv \left< \phi_{m}(t,\vartheta)
\phi_{m}(t',\vartheta')\right>$ is the SUSY correlator in
the site representation.
Averaging over  quenched disorder in intersite couplings results in
the effective  attractive interaction \cite{5}

\begin{equation}
W=-{1\over 2} \sum_{lm} w_{lm} \phi_{l}(t,\vartheta)
 \phi_{l}(t',\vartheta') \phi_{m}(t,\vartheta)
\phi_{m}(t',\vartheta').  \label{100} \end{equation}
By analogy with Eq.(\ref{99}) it is supposed that

\begin{equation}
W\simeq - {w\over 2}
C(t,\vartheta;t',\vartheta')\sum_{l}  \phi_{l}(t,\vartheta)
\phi_{l}(t',\vartheta'),\quad w\equiv \sum_{m} w_{lm} >0.
\label{101} \end{equation}
So, the real interaction (\ref{99}) contains both diagonal and
non--diagonal in  $\vartheta$ and $\vartheta'$
SUSY correlators, whereas the
quenched disorder averaging results in non--diagonal
expression (\ref{101}) only. Obviously, within the framework of the
replica approach such SUSY structure corresponds to the
inter--replica overlapping that is responsible for the specific
spin--glass behaviour  \cite{1}.

Apart from  the above contributions to SUSY action
(\ref{97}) it should be taken into account that the quenched disorder in
force dispersion  results in the additional interaction \cite{5}

\begin{equation}
\Delta S_0={h^2 \over 2} \sum_{l} \int
\phi_{{l}\omega} (\vartheta)\delta(\omega)\phi_{{l}\omega}(\vartheta)
{\rm d}\omega{\rm d}\vartheta. \label{102} \end{equation}
where $\omega$ is the frequency and the intensity of the quenched disorder

\begin{equation}
h^2={\overline{\left( f_{l} -\overline f\right)^2} -(\Delta \varphi)^2
\over(\Delta \varphi)^2 } \label{103}\end{equation}
characterizes the site dispersion of the force $f_{l}$ (overbar
denotes the volume average), $(\Delta\varphi)^2 \equiv
\varphi_{\omega=0}^2$ is mean--squared fluctuation of this force.
Then, the mean--field SUSY action takes, in the site--frequency
representation, the final form

\begin{equation}
S=\sum_{l} \int \lambda_{{l} \omega} (\vartheta)
{{\rm d} \omega\over 2 \pi} {\rm d}\vartheta+
\sum_{l} \int \lambda_{{l} \omega} (\vartheta,\vartheta') {{\rm
d} \omega\over 2 \pi} {\rm d}\vartheta  {\rm d} \vartheta'
\label{104} \end{equation}
with the SUSY Lagrangian

\def\theequation{\arabic{equation}{a}}\setcounter{equation}{104}
\begin{equation}
\lambda(\vartheta) \equiv{1\over 2} \phi(\vartheta) \left\{
\left[ 1+ D(\vartheta) \right]+
2\pi h^2\delta(\omega) - vS
\right\} \phi(\vartheta) +V_1(\phi(\vartheta)),
\label{105a}
\end{equation}
\def\theequation{\arabic{equation}{b}}\setcounter{equation}{104}
\begin{equation}
\lambda (\vartheta,\vartheta') \equiv
-{1\over 2}(v+w) \phi(\vartheta)
C(\vartheta,\vartheta') \phi(\vartheta')
\label{105b}\end{equation}
\def\theequation{\arabic{equation}}\setcounter{equation}{105}
\vspace{-0.3cm}\noindent\\ where indexes $l$, $\omega$ are suppressed for
brevity and the generator $D$ is given by Eq.(\ref{14}).

In the important case of random heteropolymer the
interaction kernels are appeared to be of the form (\ref{98}), (\ref{100}),
but indexes ${l}$, ${m}$ denote  wave  vectors but site
numbers (see \cite{ED}).
So, in this case, the expressions (\ref{104}),
(105) can be modified by replacing
site indexes by
wave ones.


\subsection {SUSY Dyson equation}

The Dyson equation for the above SUSY Lagrangian is

\begin{equation}
{\bf C}^{-1}={\bf L} - {\bf \Sigma} -(v+w){\bf C}.
\label{106} \end{equation} Here
 {\bf L} is defined by Eq.(78) where the first component
is


\begin{equation}
L= L_0 +vS, \qquad
L_0 = -(1+ 2\pi h^2\delta(\omega)).\label{107}
\end{equation} Projecting Eq.(\ref{106}) along the ''axes''
(\ref{69}), we come to the key equations
 written in the frequency
representation

\def\theequation{\arabic{equation}{a}}\setcounter{equation}{107}
\begin{equation}
S={(\Sigma-L_0) G_+G_-\over
1-wG_+G_-},\label{108a}
\end{equation}
\def\theequation{\arabic{equation}{b}}\setcounter{equation}{107}
\begin{equation}
G_\pm^{-1}+(v+w)G_\pm=L_\pm-\Sigma_\pm\label{108b}
\end{equation}
\def\theequation{\arabic{equation}}\setcounter{equation}{108}
\vspace{-0.3cm}\noindent\\ where Eq.(\ref{70}) is used.
These equations accompanied by Eqs.(96)
for the components $\Sigma$, $\Sigma_\pm$  of the SUSY self--energy
function form the closed system of equations for self--consistent
analysis of non--equilibrium thermodynamic system.

\section {Non--ergodicity and memory effects}

As it is well--known, the memory is characterized by the
Edwards--Anderson parameter \cite{EA}

\begin{equation}
q=\left<\eta (\infty)\eta(0)\right>\label{109}
\end{equation} that being  late--time
asymptotics of the correlator results in elongation of
the structure factor:

\begin{equation}
S(t)=q+S_0(t)\label{110}\end{equation}
where the component $S_0(t) \to 0$ at $t\to\infty$.
By analogy with the
elongated structure factor (\ref{110}), the ergodicity breaking
is allowed for by
adding the term to the retarded Green function

\begin{equation}
G_-(\omega)=\Delta+G_{-0}(\omega).\label{111}
\end{equation}
The non--ergodicity
parameter (irreversible response) in Eq.(\ref{111})

\begin{equation}
\Delta=\chi_0-\chi\label{112}
\end{equation}
is determined by  the adiabatic Cubo susceptibility  $
\chi _ 0\equiv G _ - (\omega = 0) $ and the  thermodynamic one $
\chi\equiv G _ {-0} (\omega =0) $.\footnote { It is convenient to use
the unique response  function $ G _ - (\omega) $ for definition
of both susceptibilities $ \chi _ 0 $ and $ \chi $, taking into
account that quantities $ \chi _ 0\equiv G _ - (\omega = 0) $, and
$ \chi\equiv G _ - (\omega\to 0) $ correspond to the equilibrium
(macroscopic) and non--equilibrium (microscopic) states.  Then
Eqs.(\ref{65}), (\ref{89}), (90), (96), where the
correlators should be labeled by index 0, imply the
limit $ \omega\to 0 $ instead of the exact equality $ \omega = 0 $.} If the
latter is defined by the standard formula $\chi={\delta
\left<\eta \right>/\delta f_{ext}}$ with
external force $f_{ext}\equiv -f$,
for determination of the former one has to use the
correlation techniques discussed in Sect.IV.

To do this let us insert the elongated correlators (\ref
{110}), (\ref {111}) into expressions (96). Then
the renormalized components of the self--energy function take the
form
\def\theequation{\arabic{equation}{a}}\setcounter{equation}{112}
\begin{eqnarray}
&&\Sigma(t)={1\over 2} \left(
\mu^2+{\lambda^2\over 3}q\right)q^2+ \left(\mu^2+
{\lambda^2\over 2}q\right)qS_0(t) +\Sigma_0(t), \nonumber\\
&&\Sigma_0(t)\equiv
 {1\over 2}\left(\mu^2 + \lambda^2q\right)S_0^2(t) + {\lambda^2
\over 6} S^3_0(t);  \label{113a}
\end{eqnarray}
\vspace{-1cm}
\def\theequation{\arabic{equation}{b}}\setcounter{equation}{112}
\begin{eqnarray}
&&\Sigma_\pm(t)=\left(
\mu^2+{\lambda^2\over 2}q\right)q\left(\Delta+G_{\pm 0}(t)\right)
+\Sigma_{\pm 0}(t),\nonumber\\
&&\Sigma_{\pm 0}(t)\equiv \left(\mu^2+\lambda^2q
\right)S_0(t)G_{\pm 0} (t) +{\lambda^2\over 2}S_0^2(t)G_{\pm 0}(t),
\label{113b}
\end{eqnarray}
\def\theequation{\arabic{equation}}\setcounter{equation}{113}
\vspace{-0.3cm}\noindent\\ where
 $ \Sigma _ 0 $, $ \Sigma _ {\pm 0} $ consist of
 the terms nonlinear in  correlators $S_0$, $G_{\pm 0}$
and the terms proportional to
$ S _ 0\Delta\simeq 0 $ are disregarded.


In the
$\omega $--representation,
inserting the  Fourier--transform of Eqs.(\ref {110}),
(\ref {113a}) in the Dyson equation  (\ref {108a}),
and taking into account Eq.(\ref {107}) we have

\begin{eqnarray}
&&q_0\left[1-w\chi_0^2 - {1\over 2}\left(\mu^2 + {\lambda^2\over
3}q_0\right)q_0\chi_0^2\right] =h^2\chi_0^2,
\label{114}\\
&&\qquad S_0={(1+\Sigma_0)G_+G_-\over
1-\left[w+(\mu^2+\lambda^2q/2)q\right]G_+G_-}.\label{115}
\end{eqnarray}
The first of these equations corresponds  to
$ \delta $--terms ($\omega=0$) that are caused by  the memory
effects, whereas the second one --- to  non--zero frequencies $
\omega\ne 0 $.  In the limit $ \omega\to 0 $, the
product $ G_+G_-$ tends to $\chi ^ 2 $,
so that the  pole of the structure factor
(\ref {115}) determines the point of ergodicity breaking for the
thermodynamic system

\begin{equation}
\chi_0^{-2}=w+\left(\mu^2+{\lambda^2\over 2}q_0\right)q_0.
\label{116} \end{equation}

Substituting Eq.(\ref{113b}) into
the Dyson equation (\ref{108b})
yields the relation
for
retarded Green function

\begin{equation}
G_-^{-1}+\left[(v+ w)+\left( \mu^2+{\lambda^2\over 2}q\right)q \right]
G_-+\Sigma_{-0} -(1-{\rm i}\omega)=0
\label{117}\end{equation}
where the $\omega$--representation is used.
Then, from Eq.(\ref{96c}) the equation for the
thermodynamic susceptibility $\chi\equiv G_-(\omega\to 0)$
is derived

\begin{equation}
1-\chi+(v+w)\chi^2+{\mu^2\over
2}\chi\left[(\chi+q)^2-q^2\right]+{\lambda^2\over
6}\chi\left[(\chi+q)^3-q^3\right]=0.
\label{118}\end{equation}
The macroscopic memory parameter $q_0$ is given by
the equation
\begin{equation}
\left( {\mu^2\over 2 } + {\lambda^2 \over 3}q_0\right)q_0^2=h^2,
\label{119}\end{equation}
which is obtained from Eqs.(\ref{114}), (\ref{116}) in the limit
$\omega= 0$.
\section{Discussion}

Within the framework of the model  under consideration
the system of Eqs.(\ref{114}), (\ref{118}), (\ref{116}), (\ref{119}),
(\ref{112}) provides the complete analytical description of the
non--ergodic thermodynamic system with quenched disorder.
Eqs.(\ref{114}) and (\ref{118})
 are similar to the equations obtained by
Sherrington and Kirkpatrick for determination of isothermal
$\chi _ 0 $ and thermodynamic $ \chi $ susceptibilities and
corresponding memory parameters $q_0$, $q$ \cite{1}.  The equation
(\ref {116}) defines the point $T_0 $ of ergodicity breaking  and
Eq.(\ref{112}) gives non-ergodicity parameter $ \Delta $.
The above consideration implies that one should distinguish the macroscopic
quantity $q_0$, $\chi_0$ and microscopic ones $q$, $\chi$ (the
former correspond to frequency $\omega=0$, the latter --- to the limit
$\omega\to 0$). The peculiarity of such a hierarchy is that
macroscopic values $q_0$, $\chi_0$ depend on the amplitude $h$ of
quenched disorder only, whereas microscopic ones $q$, $\chi$ --- on
temperature $T$.  Respectively,  Eqs. (\ref{114}), (\ref{116}), where
the temperature $T$ should be taken
equal to its value  on the ergodicity breaking curve $T_0(h)$, give
the macroscopic values $q_0$, $\chi_0$.
What about determination of
the microscopic ones $q$, $\chi$, the Eq.(\ref{118}) must
be added by the equation type of Eq.(\ref{114})
 \begin{eqnarray}
&&q\left[1-w\chi_0^2 - {1\over 2}\left(\mu^2 + {\lambda^2\over
3}q\right)q\chi_0^2\right] =h^2\chi_0^2,
\label{120}
\end{eqnarray}
where the memory parameter $q$ is taken as microscopic in character.

It is well to bear in mind that  the
 field $ h $, anharmonicity parameters $\lambda$, $\mu$,
and  interaction parameter $ w $,
as well as the inverse values of susceptibilities $ \chi _ 0 $,
$ \chi $ have been measured in units of temperature $ T $.
Further, it is  convenient to choose the following measure
units:

\begin{eqnarray}
&&T_s=\left({3\over 2}\right)^{3/2} {\mu^4\over \lambda^3}, \
h_s={3\over 2}\, {\mu^3\over \lambda^2},\
v_s=w_s=\left({3\over 2} \right)^{-1/2}\lambda, \nonumber\\
&&\chi_s = \left({3\over 2} \right)^{-1/2}{\lambda\over \mu^2},\
q_s = {3\over 2} {\mu^2\over \lambda^2}\equiv u \label{121}
\end{eqnarray}
for quantities $T$, $h$, $v$, $w$, $\chi$, $q$
respectively.  As a result, the key equations take the final form:

\def\theequation{\arabic{equation}{a}}\setcounter{equation}{121}
\begin{eqnarray}
&&\left(1- u T\chi\right)+(v+w)T\chi^2+ (\chi/
2T)\left[(T\chi+q)^2-q^2\right]+\nonumber\\
&&\qquad\qquad\qquad\qquad\qquad\qquad\qquad\qquad
(\chi/ 4T)\left[(T\chi+q)^3-q^3\right]=0, \label{122a}
\end{eqnarray}
\def\theequation{\arabic{equation}{b}}\setcounter{equation}{121}
\begin{equation}
w T+\left( 1+q/2\right)q/2+ h^2/q=\chi_0^{-2},\label{122b}
\end{equation}
\def\theequation{\arabic{equation}{c}}\setcounter{equation}{121}
\begin{equation}
w T+\left( 1+q_0/2\right)q_0/2+ h^2/q_0=\chi_0^{-2},\label{122c}
\end{equation}
\def\theequation{\arabic{equation}{d}}\setcounter{equation}{121}
\begin{equation}
\left[1 +(3/4)q_0\right]q_0+ wT_0=\chi_0^{-2},
\label{122d}
\end{equation}
\def\theequation{\arabic{equation}{e}}\setcounter{equation}{121}
\begin{equation}
(1+q_0)q_0^2=2h^2,\label{122e}
\end{equation}
\def\theequation{\arabic{equation}{f}}\setcounter{equation}{121}
\begin{equation}
\Delta= u T(\chi_0-\chi).\label{122f}
\end{equation}
\def\theequation{\arabic{equation}}\setcounter{equation}{122}
\vspace{-0.3cm}\noindent\\
Behaviour of the system is specified by the last parameter
$u$ in Eqs.(121), that determines the relation
between cubic and quartic anharmonicities.
In the case $u\ll 1$ main
contribution is due to the quartic term (\ref{81}),
whereas at $u\gg 1$ the cubic  anharmonicity
(\ref{91}) dominates.
The former limit corresponds to strong quenched disorder
$h\gg h_s$,  the latter~--- to weak field $h\ll h_s$.

According to  Eq.(\ref{122e}), the dependence of the macroscopic
memory parameter $q_0$ on the quenched disorder amplitude
$h$ is governed by the ratio between magnitude $h$ and
characteristic field $h_s$.
The linear dependence

\begin{equation}
q_0=2^{1/2}(h/\mu),\quad u\gg 1, \label{123} \end{equation}
is realized in the limit $h\ll h_s$, whereas the
power relation

\begin{equation}
q_0=3^{1/3}(h/\lambda)^{2/3},\quad u\ll 1\label{124}
\end{equation}
corresponds to the case of $h\gg h_s$
(in Eqs.(\ref{123}), (\ref{124}) measured units are restored).
The dependence $q_0(h)$ is depicted in Fig.1.

For temperatures above the point  of  ergodicity
breaking $T_0$ the thermodynamic $\chi$, $q$ and
adiabatic $\chi_0$, $q_0$ values
of susceptibilities and memory parameters, as well as Eqs.(122b), (122c)
coincide, so that Eqs.(122a),(122c)
deteremine the dependencies $\chi$ and $q$ on temperature.
Accounting  Eq.(122d) gives the  temperature
of the ergodicity breaking $T_0(h)$ as a function of field $h$
(see Fig.2).
The  peculiarity of the dependence  $T_0(h)$ is that the
ergodicity breaking temperature takes non--zeroth value
$T_{00}\equiv T_0(h=0)$ at $h=0$.
Below the ergodicity breaking temperature $T_0$
Eqs.(\ref{122a}), (\ref{122b}) give the
microscopic values of the memory parameter $q$
and the susceptibility $\chi$ that differs from
the macroscopic one $\chi_0$ being constant.
As a result, we obtaine  the typical temperature dependencies
of susceptibilities $\chi$, $\chi_0$ as shown in Fig.3.
It is seen that, in accordance with Eq.(122a), at $T<T_0$
the thermodynamic susceptibility  $\chi\ne 0$
only if temperature is above the freezing point $T_f$ which is
determined  by the
condition $\partial \chi/\partial T=\infty$, leading to the
equation

\begin{equation}
(v+w)T_f+ T_f\chi+q+(3/ 4)(T_f\chi+q)^2=\chi^{-2},
\label{125}\end{equation}
where $\chi$, $q$ are taken at $T=T_f$.

The phase diagrams, which depict the ranges of
possible thermodynamic states on the plane $h-T$
for various values of interaction parameters $w$ and $v$,
are shown in
Fig.2.  In the limit $h=0$ the  temperatures of
ergodicity breaking and freezing are as follows:

\def\theequation{\arabic{equation}{a}}\setcounter{equation}{125}
\begin{equation}
T_{00}=w\left\{ \left( 1+ {v\over 2w}+ {1\over 12}{\lambda^2\over
w^2} \right) +\left[ \left( 1 + {v\over 2w}+ {1\over 12}{
\lambda^2\over w^2} \right)^2 +{1\over 2}{\mu^2\over w^2} \right]
^{1/2}\right\}^2,\label{126a}
\end{equation}
\def\theequation{\arabic{equation}{b}}\setcounter{equation}{125}
\begin{equation}
T_f \approx 4(v+w)\left(1+
{\mu^2 +(2/3) \lambda^2 \over 4(v+w)^2}
\right) , \label{126b} \end{equation}
\def\theequation{\arabic{equation}}\setcounter{equation}{126}
\vspace{-0.3cm}\noindent\\ where measured units are restored and
the second equality is for $\mu^2, \lambda^2 \ll w^2$.
So, cubic and quartic  anharmonicities result in an increase of
both ergodicity breaking and freezing temperatures.
When quenched disorder is large, $q_0, q_0^2\gg  wT_0$,
the isothermal susceptibility $\chi_0$ in Eq.(\ref{122d}) is
small, and Eqs.(122a), (122c), (122d) provide the estimate
$\chi_0\approx 2/uT_0$.
Then,  for measured quantities one has:

\def\theequation{\arabic{equation}{a}}\setcounter{equation}{126}
\begin{equation}
T_0\approx 2^{5/4}\ \mu\ ( h/\mu)^{1/2}, \qquad (w/\mu)^2\mu  \ll
h\ll (\mu/\lambda)^2\mu; \label{127a}
\end{equation}
\def\theequation{\arabic{equation}{b}}\setcounter{equation}{126}
\begin{equation}
T_0\approx
2^{1/2} 3^{1/3}\ \lambda\ (h/\lambda)^{2/3}, \qquad  h\gg
(\mu/\lambda)^2\mu,\  (w/\lambda)^{3/2}\ \lambda
\label{127b} \end{equation}
\def\theequation{\arabic{equation}}\setcounter{equation}{127}
\vspace{-0.3cm}\noindent\\ for $\mu^2\gg \lambda^2$
and $\lambda^2\gg \mu^2$, respectively.
So, the non--ergodicity domain is extended indefinitely at strong
increasing of quenched disorder.
As it is shown in Figs.2a,b,
and  Fig.4a,b, the dependencies $T_0(h)$, $T_f(h)$ is non--monotonous
if either $w< 0.5$ or $v>1$.

Influence of the interaction parameters
$w$, $v$ and the anharmonicity ratio $u$ on
the temperature dependence
of the susceptibility $\chi$ is illustrated in Fig.5.  According to
Fig.5a, increasing $w$ causes a decrease of $\chi$ and
an increase of temperatures $T_0$, $T_f$. The same
behaviour is revealed at increasing parameter $v$ (Fig.5b). By
contrast, tendency is opposite under
an increase in $u$ (see  Fig.5c).

Finally, let us consider behaviour of the   non--ergodicity $\Delta$
and memory $q$ parameters that are determined by complete system
of equations (122).
Corresponding dependencies on temperature are shown in Fig.6a,b.
At the freezing state,
where $ \chi\equiv 0 $, the non--ergodicity parameter (\ref {122f})
linearly depends on temperature because the isothermal
susceptibility $\chi_0$ is constant.  The appearance of finite value of
the thermodynamic susceptibility $ \chi $ above the freezing point
$T_f$ results in step--like decrease of the value $ \Delta $.
With further growth of temperature the irreversible
response $\Delta(T)$ monotonously decays taking zero value
at the ergodicity breaking point $T_0$ (see Fig.6a).
With increasing temperature from 0 to $T_0$
microscopic memory parameter $q$
monotonously decriases, taking minimal value at the ergodicity
breaking point $T_0$.
Above this point  $q(T)$ increases (see Fig.6b).
It is seen, that the quenched disorder encrease extendes the
temperature domain of the non--ergodicity and causes  growth of the memory
parameter.
In a spirit of generalized picture of phase
transition it can be attributed to
the fact that the microscopic memory parameter $q$ above the point $T_0$
corresponds to a soft mode that transforms to a mode of ergodicity
restoring below the temperature $T_0$. The non--ergodicity
parameter $\Delta$ represents the order parameter.

The analytical expressions for dependencies $ \Delta (T) $, $q(T)$
can be obtained only near the  ergodicity breaking curve $T_{0}(h)$.
For $h=0$ and $T_0 = T_{00}$, from  Eq.(\ref {122a})
assuming that $ 0 < T_{00}-T\ll T_{00}$,
$\chi\approx \chi_{00} -\Delta/uT_{00} $,
$ \Delta\ll u\chi_{00}T_{00} $, up to the first order in small parameters
$ \varepsilon \equiv T/T_{00}-1 $ and
$ \Delta (u\chi_{00} T_{00})^{-1}$  we have for measured units

\def\theequation{\arabic{equation}{a}}\setcounter{equation}{127}
\begin{equation}
\Delta=-A_0\varepsilon,\quad
A_0\equiv {T_{00}\over w} \left({w\over
\mu}\right)^2 {1 - {\lambda^2\over 6w^2}\over
1+\left({\lambda^2\over 2\mu^2}+ {vw\over\mu^2}\right)
\left({T_{00}\over w}\right)^{1/2}},\quad \varepsilon<0;
\label{128a}
\end{equation}
\vspace{-0.5cm}
\def\theequation{\arabic{equation}{b}}\setcounter{equation}{127}
\begin{equation}
q= Q\varepsilon,\quad Q \equiv {4\over
3} {T_{00}\over w} \left({\lambda w\over \mu^2}\right)^2
{1 - {\lambda^2\over 12w^2}\over  1+{\lambda^2\over 2\mu^2}
\left({T_{00}\over w}\right)^{1/2}},\quad \varepsilon>0,
\label{128b} \end{equation}
\def\theequation{\arabic{equation}}\setcounter{equation}{128}
\vspace{-0.3cm}\noindent\\
In the case of $h\ne 0$  the result for temperature dependence is

\begin{equation}
\Delta=-A\varepsilon,\quad A\equiv {2\over \lambda^2 \chi_0^2}\left(
{ 1-{w\over 2}\chi_0^2 T_0-{\lambda^2\over 12} \chi_0^4T_0^2 \over
{v\over \lambda^2\chi_0}+ {\mu^2\over \lambda^2}+q_0+
{1\over 2}\chi_0 T_0}\right), \quad \Delta,\ \varepsilon<0.
\label{129} \end{equation}
Correspondingly, at the fixed temperature Eq.(\ref{122a})  gives
in the linear approximation $0< q_0-q \ll q_0$:

\begin{equation} \Delta=B \left(q- q_0
\right),\quad B^{-1}\equiv 1+ {v\over \lambda^2 \chi_0}\left(
{\mu^2\over \lambda^2}+q_0+ {1\over 2}\chi_0 T_0\right)^{-1}.
\label{130} \end{equation}

\newpage
\appendix

\section{}

For nilpotent representation  let us rewrite the Lagrangian
(\ref{10}) in the form of Euclidean field theory \cite{8}

\begin {equation}
L = \kappa + \pi, \label {A.1} \end{equation}
where the kinetic $\kappa$ and potential $ \pi  $ energies are

\begin {eqnarray}
&& \kappa = \varphi\dot\eta -{\varphi ^2/ 2},
\label {A.2} \\
&& \pi =   (\partial V/ \partial \eta)\varphi.
\label {A.3}
\end{eqnarray}
In order to obtain the nilpotent
form (\ref {13a}) of the kinetic energy  (\ref{A.2}),
we have to determine the operator $ D$.
The complete form of the dependence of the operator $D$ on the
nilpotent coordinate $ \vartheta $ is presented by the expression

\begin {equation}
D = a + b ({\partial/ {\partial \vartheta}}) + c\vartheta + d\vartheta ({\partial/
{\partial \vartheta}}), \label {A.4} \end {equation}
where the coefficients
$ a $, $ b $, $ c $, $ d $ are unknown operators.
The  substitution of Eqs.(\ref {11}), (\ref {A.4})
into Eq.(\ref {13a}) and taking into account the properties (\ref {12})
leads  to the expression (\ref {A.2}) with the following coefficients:

\begin {equation} a = \partial_t, \qquad b = -1,
\qquad c = 0, \qquad d = -2\partial_t, \label {A.5}
\end {equation}
where $\partial_t\equiv \partial /\partial t$ is the derivative with
respect to time.
As a result, the operator (\ref {A.4}) takes the form (\ref{14}).
It has the property

\begin {equation}
D^{2} = \partial^{2}_t.  \label {A.6} \end {equation}
While considering the definitions (\ref {11}), (\ref{12}),
(\ref {14}) it is easy to see that  $ D $  is a Hermite operator.

Under the infinitesimal transformation
$\delta\equiv e ^ {\varepsilon D} -1 \simeq \varepsilon D $
with the parameter $ \varepsilon\to 0 $, the  values $ t $ and
$ \vartheta $  acquire the additions $ \delta t = \varepsilon$,
$\delta \vartheta = -\varepsilon$ which differ in the sign.
Considering  the corresponding field addition
$|\delta\phi_\varphi|=\varepsilon |D_\varphi| |\phi_\varphi|$,
it is convenient to use the matrix form  for
the nilpotent field (\ref{11}) and the operator $D$:

\begin {equation}
|\phi_\varphi|= {\eta \choose \varphi},\qquad |D_\varphi|=\left(
\begin{array}{cc} \partial_t & -1\\
0 & -\partial_t \end {array}
\right),\qquad  \partial_t\equiv {\partial\over\partial
t}.  \label {A.7} \end {equation}
According to
Eq.(\ref{A.7}), the change of the order parameter is proportional to a
difference between the rate of change of the order parameter and
the fluctuation amplitude, whereas the  change of the latter is
proportional to its rate with the opposite sign.

To prove the equivalence of the term (\ref{A.3}) in the Lagrangian
(\ref{A.1}) and the Grassmann potential energy (\ref {13b}), let us carry
out the formal expansion of the thermodynamic potential in powers of the
component $\vartheta\varphi$ of Eq.(\ref{11}):

\begin {equation}
\pi=\int \left[ V(\eta) +{\delta V\over \delta \eta}\
\varphi \vartheta \right] {\rm d}\vartheta. \label {A.8} \end {equation}
Here all the terms of powers higher than 1 are omitted according to the
nilpotent condition. Using the integration properties (\ref{12}), we
obtain immediately Eq.(\ref{A.3}) as it was required.

In the case of  the two--component nilpotent
field (\ref{27}), the consideration is
fulfilled by analogy. For brevity, let us point out the
difference between (\ref{27}) and the above--considered case (\ref{11}) only.
The corresponding infinitesimal transformation
$\delta \simeq \varepsilon D$, where the generator $D$ is given by Eq.(\ref{28}),
results in the additions $ \delta t = 0$,
$\delta \vartheta = -\varepsilon$,
$|\delta{\bf {\phi}}_f|=\varepsilon |D_f| |\phi_f|$
in which the matrix form

\begin {equation}
|\phi_f|=\left( {\eta \atop -f}\right), \qquad
|D_f| = \left( \begin{array} {cc} 0 & -1 \\
-\partial^2_t & 0 \end{array}\right),\qquad \partial_t\equiv
{\partial\over \partial t} \label {A.9} \end {equation}
is used.
The generator $D_f$ has the
property (\ref{A.6}).  It is easy seen that the Lagrangians
(\ref{10}),  (\ref{23}) are invariant under the
transformations given by the generators $D_\varphi$, $D_f$,
respectively, provided that the infinitesimal parameter
$\varepsilon$ is pure imaginary, and the fields $\eta({\bf r}, t)$,
$\varphi({\bf r}, t)$, $f({\bf r}, t)$ are complex--valued.  So, for
real fields the two--component representations (\ref{11}), (\ref{27})
are just convenient approximations.

The matrices of the transformation between the fields (\ref{A.9}) and
(\ref{A.7}) (see Eqs.(\ref{29}), (\ref{32}))
take the form

\begin{equation} |\tau_\pm|=\left(
\begin{array}{cc} 1 & 0\\ \pm\partial_t & 1 \end{array}
\right).\label{A.10} \end{equation}

Let us consider the four--component SUSY fields (\ref{35}),
(\ref{44}). Instead of Eq.(\ref{A.6}),
the corresponding couples of operators (\ref{37}),
(\ref{45})  satisfy the conditions:

\begin{eqnarray}
&&{\cal D}^2=\overline{\cal D}^2=0,\quad
\{\overline {\cal D},{\cal D}\}=-2\partial_t,\quad [\overline {\cal
D},{\cal D}]^2=(2\partial_t)^2;\label{A.11}\\
&&\{ {\cal D}_\varphi,{\cal D}_f\}=\{\overline {\cal D}_\varphi,
\overline{\cal D}_f\}=0,\quad
\{\overline {\cal D}_\varphi,{\cal D}_f\}=-\partial_t,\quad
\{\overline {\cal D}_f, {\cal D}_\varphi\}=-3\partial _t,
\nonumber\end{eqnarray}
where the curly and square brackets denote anticommutator
and commutator, respectively.
The generalized anticommutation rules for the operators
${\cal D}^{(\pm)}\equiv {\cal D} (\pm t)$,
$\overline{\cal D}^{(\pm)}\equiv \overline {\cal D} (\pm t)$
corresponding to the opposite directions of the time $t$, read:

\begin{eqnarray}
&&\{ {\cal D}^{(\pm)},{\cal D}^{(\mp)}\}=
\{\overline {\cal
D}^{(\pm)},\overline {\cal D}^{(\mp)}\}=\{\overline {\cal
D}^{(\pm)}_f,{\cal D}^{(\mp)}_f\}=0,\nonumber\\
&&\{\overline {\cal
D}^{(\pm)},{\cal D}^{(\pm)}\}=-\{\overline {\cal
D}^{(\pm)}_\varphi,{\cal D}^{(\mp)}_\varphi\}=\mp 2\partial_t;\nonumber\\
&&\{ {\cal D}^{(\pm)}_\varphi,{\cal D}^{(\pm)}_f\}=
\{\overline {\cal
D}^{(\pm)}_\varphi,\overline {\cal D}^{(\pm)}_f\}=0,
\quad\{\overline {\cal D}^{(\pm)}_\varphi, {\cal
D}^{(\pm)}_f\}=\mp\partial_t,\quad  \{\overline {\cal
D}^{(\pm)}_f ,{\cal D}^{(\pm)}_\varphi\}=\mp 3\partial_t;
\nonumber\end{eqnarray}
\begin{eqnarray}
&&\{{\cal D}^{(\pm)}_\varphi, {\cal D}^{(\mp)}_f\}=
\{\overline {\cal D}^{(\pm)}_\varphi, \overline{\cal D}^{(\mp)}_f\}=0,\qquad
\{\overline {\cal D}^{(\pm)}_\varphi, {\cal D}^{(\mp)}_f\}=
\{\overline {\cal D}^{(\pm)}_f ,{\cal
D}^{(\mp)}_\varphi\}=\pm\partial_t.
\label{A.12} \end{eqnarray}
In Eqs.(\ref{A.11}), (\ref{A.12}) the coincident
indexes are suppressed. The simplest way to prove Eqs.(\ref{A.11}),
(\ref{A.12}) is to introduce the four--rank matrices
(see Eqs.(\ref{A.7}), (\ref{A.9})):

\begin{eqnarray}
&& |\Phi_\varphi| =
\left(\begin{array} {c} \eta\\ \psi\\ -\overline\psi\\ \varphi
\end{array}  \right), \quad
|{\cal D}_\varphi|=
\left(\begin{array} {cccc}
0&1&0&0\\
0&0&0&0\\
-2\partial_t &0&0&1\\
0&2\partial_t&0&0
\end{array}  \right),
 \quad
|\overline {\cal D}_\varphi|=
\left(\begin{array} {cccc}
0&0&1&0\\
0&0&0&-1\\
0&0&0&0\\
0&0&0&0
\end{array}  \right);\label{A.13}\\
&& |\Phi_f| =
\left(\begin{array} {c} \eta\\ \psi\\ -\overline\psi\\ -f
\end{array}  \right), \quad
|{\cal D}_f|=
\left(\begin{array} {cccc}
0&1&0&0\\
0&0&0&0\\
-\partial_t &0&0&1\\
0&\partial_t&0&0
\end{array}  \right),
 \quad
|\overline {\cal D}_f|=
\left(\begin{array} {cccc}
0&0&1&0\\
-\partial_t&0&0&-1\\
0&0&0&0\\
0&0&-\partial_t&0
\end{array}  \right).  \label{A.14} \end{eqnarray}
The matrices of the transformation between the fields (\ref{A.14}) and
(\ref{A.13}) take the form (cf. Eq.(\ref{A.10}))

\begin{equation}
|T_\pm|=
\left(\begin{array} {cccc}
1&0&0&0\\
0&1&0&0\\
0&0&1&0\\
\pm\partial_t&0&0&1
\end{array}  \right).\label{A.15}
\end{equation}
The infinitesimal transformations
$\delta\simeq \overline\varepsilon {\cal D}$,
$\overline\delta\simeq \overline {\cal D}\varepsilon$ give
the following additions:

\begin{eqnarray} &&
\delta_\varphi\theta =0,\quad
\delta_\varphi\overline\theta=\overline\varepsilon,\quad \delta_\varphi
t=-2\overline\varepsilon\theta,\quad
\overline\delta_\varphi\theta=\varepsilon,\quad
\overline\delta_\varphi \overline\theta=0,\quad \overline\delta_\varphi
t=0;\nonumber \\
&&\delta_f\theta =0,\quad
\delta_f\overline\theta=\overline\varepsilon,\quad \delta_f
t=-\overline\varepsilon\theta,\quad
\overline\delta_f\theta=\varepsilon,\quad
\overline\delta_f \overline\theta=0,\quad \overline\delta_f
t=-\overline\theta\varepsilon;\label{A.16}
\end{eqnarray}
\begin{eqnarray}
&& |\delta\Phi_\varphi |=\overline\varepsilon |{\cal D}_\varphi
||\Phi_\varphi|, \quad |\overline\delta\Phi_\varphi |=
|\overline{\cal D}_\varphi ||\Phi_\varphi| \varepsilon; \nonumber\\
&& |\delta\Phi_f |=\overline\varepsilon |{\cal D}_f ||\Phi_f|, \quad
 |\overline\delta\Phi_f |= |\overline{\cal D}_f ||\Phi_f|
\varepsilon.\nonumber
\end{eqnarray}

At last, the equation for the SUSY field (\ref{35})

\begin{equation}
\int V(\Phi_\varphi){\rm d}^2 \theta = {\delta
V\over \delta\eta}\ \varphi -\overline\psi\ {\delta^2 V\over
\delta\eta^2}\ \psi \label{A.17}
\end{equation}
is obtained by analogy with Eq.(\ref{A.8}) to represent
the terms in Eq.(\ref{34}) that contain the potential $V\{\eta\}$ in SUSY form.
In the case of the field (\ref{44}), the multiplier $\varphi$ must be
substituted by $-f$.

\section{}

Following the standard field scheme \cite {4}, let us show how
the four-component SUSY field (\ref{44}) is split into a couple
of chiral two--component Grassmann conjugated fields
$ \Phi _\pm $. These SUSY fields are obtained
from the initial SUSY field $ \Phi_f$ under the following transformations:

\begin {equation}
\Phi _ \pm = T_\pm \Phi_f;
\quad T _ \pm\equiv e ^ {\pm\partial}, \quad \partial
\equiv  \overline\theta\theta \partial_t, \qquad
\partial_t\equiv \partial/\partial t. \label {B.1}
\end {equation}
Accordingly, the generators (\ref{45})  take the form

\begin {equation}
{\cal D}_\pm = T_\pm {\cal D}_fT_\mp \qquad \overline{\cal D}_\pm =
T_\pm\overline{\cal D}_f T_\mp.  \label {B.2} \end {equation}
Due to the Grassmann nature
of the parameter $ \partial $ in the operators $ T _ \pm $, it is
convenient to rewrite (\ref {B.2}) in the following form:

\begin{equation}
{\cal D} _ \pm = {\cal D}_f\pm\left [\partial, {\cal
D}_f\right], \qquad \overline {\cal D} _ \pm = \overline {\cal
D}_f\pm \left [\partial, \overline {\cal D}_f\right], \label {B.3}
\end {equation}
where the square brackets denote commutator. In explicit form one has

\begin {eqnarray}
&& {\cal D} _ + = {\partial/ \partial\overline\theta} - 2\theta
\partial_t, \qquad {\cal D} _ - = {\partial/ \partial
\overline\theta}; \nonumber \\ &&\overline {\cal D} _ + = {\partial/
\partial \theta}, \qquad \overline {\cal D} _ - = {\partial/
\partial\theta} - 2\overline\theta \partial_t.
\label{B.4} \end {eqnarray}
Apparently, the operators ${\cal D}_+$, $\overline{\cal D}_+$
coincide with the generators ${\cal D}_\varphi$,
$\overline{\cal D}_\varphi$, Eqs.(\ref{37}).

According to Eqs.(\ref{44}), the definitions (\ref {B.1}) give

\begin{equation}
\Phi _ \pm = \eta + \overline\theta\psi + \overline\psi\theta \pm
\overline\theta\theta\left (\dot\eta\mp f\right), \label
{B.5} \end {equation}
where the point denotes the  derivative with respect to time. The comparison of
Eq.(\ref{B.5}) with the definition (\ref{35}) gives the identity
$\Phi_+\equiv \Phi_\varphi$.
The action of the operators (\ref {B.4}) on Eq.(\ref{B.5}) gets

\begin {eqnarray}
&& {\cal D} _ \pm\Phi _ \pm = \psi - \theta\left (
 \dot\eta+f\right) + \underline {2\overline\theta\theta\dot\psi},
\nonumber \\
&& -\overline {\cal D} _ \mp\Phi _ \mp = \overline\psi +
\overline\theta\left ( \dot\eta-f\right) - \underline
{2{\overline\theta} \theta\dot {\overline\psi}}, \label {B.6} \end
{eqnarray}
where the underlined terms concern only  the upper indexes of
the left--hand parts.

The chiral SUSY fields are fixed by the gauge conditions \cite {4}

\begin {equation} {\cal D} _ -\Phi _ - = 0, \qquad
\overline {\cal D} _ + \Phi _ + = 0, \label {B.7} \end {equation}
which, in accordance with the definitions (\ref {B.4}) signify, that $ \Phi _ - $
and $ \Phi _ + $ are independent of  $ \overline\theta $ and
$ \theta $, respectively.
On the  other hand, taking into account Eqs.(\ref {B.6}), the
gauge (\ref {B.7}) results in the equations

\begin {eqnarray}
&&\psi-\theta\left(f+ \dot\eta\right)=0,\nonumber\\
&&\overline\psi+\overline\theta\left(\dot\eta-f\right)=0
\label {B.8} \end {eqnarray}
for $\Phi_-$ and $\Phi_+$, correspondingly.
Substituting Eqs.(\ref{B.8})
into Eq.(\ref {B.5}), the final expressions for the
chiral SUSY fields are obtained:

\begin {eqnarray}
&&\phi_-=\eta+\overline\psi\theta,\nonumber\\
&&\phi_+=\eta+\overline\theta\psi.
\label {B.9} \end {eqnarray}
These equations  give the  non--reducible
representations of the SUSY fields (\ref{35}), (\ref{44})
under the conditions of the gauge (\ref{B.7}).
The chiral field $\phi_+(t)$ corresponds to the
positive direction of the time $t$, whereas $\phi_-(t)$ is related to the
negative one \cite{4}.

\section{}

Let us consider the invariance properties of the SUSY action

\begin{equation}
S=\int \Bigl[ K(\Phi(z)) + V(\Phi(z))\Bigr] {\rm d} z,\quad
K(\Phi)\equiv {1\over 2} (\overline{\cal D}\Phi) ({\cal
D} \Phi), \quad z\equiv \{{\bf r}, t, \theta,
\overline\theta\}\label{C.1} \end{equation}
under the Grassmann conjugated transformations

\begin{equation}
\delta\Phi = \sum_\alpha \overline\varepsilon_\alpha
{\cal D}^{(\alpha)}\Phi,\qquad \overline\delta\Phi = \sum_\alpha
\overline{\cal D}^{(\alpha)}\Phi\varepsilon_\alpha,
\label{C.2} \end{equation}
given by the SUSY generators ${\cal D}^{(\alpha)}$,
$\overline {\cal D}^{(\alpha)}$ which differ in
the time $t$ and Grassmann
coordinates $\theta$, $\overline\theta$.  According to
Eq.(\ref{A.17}),  the potential term in Eq.(\ref{C.1}) is SUSY
invariant if the kernel $V(\eta)$ does not depend on the time
$t$.  Up to inessential total time derivatives,  the Grassmann
conjugated variations of the remaining kinetic term

\begin{eqnarray}
&&\delta K= {1\over 2}\overline{\cal D} \left(
\sum_\alpha
\overline\varepsilon_\alpha  {\cal D}^{(\alpha)}\Phi\right) ({\cal
D}\Phi) + {1\over 2}(\overline{\cal D}\Phi) {\cal D} \left(\sum_\alpha
\overline\varepsilon_\alpha {\cal D}^{(\alpha)} \Phi\right),\label{C.3}\\
&&\overline \delta K= {1\over 2}\overline{\cal D} \left(
\sum_\alpha
\overline  {\cal D}^{(\alpha)}\Phi \varepsilon_\alpha\right) ({\cal
D}\Phi) + {1\over 2}(\overline{\cal D}\Phi) {\cal D} \left(\sum_\alpha
\overline {\cal D}^{(\alpha)} \Phi \varepsilon_\alpha\right)
\nonumber \end{eqnarray}
can be rewritten in the form

\begin{equation}
\delta K ={1\over 2}
\sum_\alpha\overline\varepsilon_\alpha {\cal D}^{(\alpha)}
\left[(\overline{\cal D}\Phi)({\cal D}\Phi)\right],\qquad
\overline\delta K ={1\over 2}
\sum_\alpha \overline{\cal D}^{(\alpha)}
\left[(\overline{\cal D}\Phi)({\cal D}\Phi)\right]\varepsilon_\alpha
\label{C.4} \end{equation}
provided that the anticommutators
$\{{\cal D}, {\cal D}^{(\alpha)}\}$,
$\{\overline{\cal D}, {\cal D}^{(\alpha)}\}$,
$\{{\cal D}, \overline{\cal D}^{(\alpha)}\}$,
$\{\overline{\cal D}, \overline{\cal D}^{(\alpha)}\}$
are either equal to zero  or  proportional to the derivative with respect to the time
$\partial_t$. According to Eqs.(\ref{A.11}), (\ref{A.12}) such conditions
are fulfilled if only the SUSY generators ${\cal D}^{(\alpha)}$,
$\overline {\cal D}^{(\alpha)}$ either coincide with the initial operators
${\cal D}$, $\overline{\cal D}$, or are reduced to the transformed operators
${\cal D}_\pm$, $\overline{\cal D}_\pm$ determined by equations
of (\ref{B.2}) type, or correspond to the opposite time directions
${\cal D}^{(\pm)}_\pm$, $\overline{\cal D}^{(\pm)}_\pm$.
Being  reduced to the derivatives with respect to the time
$t$ and Grassmann coordinate $\theta$, $\overline\theta$,
these  operators inserted into Eqs.(\ref{C.4}) give, as it
was required, zero for the variations of the corresponding action  (\ref{C.1}).

Among the above--mentioned generators,  the following ones

\begin{eqnarray}
&& {\cal D}_-^{(-)}={\partial \over \partial
\overline\theta}, \qquad
\overline{\cal D}_-^{(-)}={\partial \over
\partial \theta}+  2\overline\theta {\partial \over \partial t}
\label{C.5}\end{eqnarray}
and their  anticommutator
$\{\overline{\cal D}^{(-)}_-,  {\cal D}^{(-)}_-\}= 2\partial_t$
(see Eqs.(\ref{A.12})) are of a special interest for us.
The operators (\ref{C.5}) are a result of the double action of the
transformation $T_-$ on the initial generators ${\cal D}_\varphi$,
$\overline {\cal D}_\varphi$, Eqs.(\ref{37}), that gives the
generators ${\cal D}_-$, $\overline {\cal D}_-$, Eqs.(\ref{B.3}),
corresponding to the opposite time directions.
Therefore, the generators
${\cal D}^{(-)}_-\equiv {\cal D}_-(-t)$,
$\overline {\cal D}_-^{(-)}\equiv \overline{\cal D}_-(-t)$ given by Eqs.(\ref{C.5})
are related to the initial ones ${\cal D}_\varphi(t)$,
$\overline {\cal D}_\varphi (t)$ and play a significant role hereinafter.

Due to the standard manner, it is easy to show that the above
conditions $\delta S=0$, $\overline\delta S=0$ give rise to the Ward
identities \cite{8}

\begin{equation}
\sum_{i=1}^n {\cal D}_i^{(\alpha)} \Gamma^{(n)}(\{ z_i \})=0, \qquad
\sum_{i=1}^n \overline{\cal D}_i^{(\alpha)} \Gamma^{(n)}(\{ z_i \})=0
\label{C.6}\end{equation}
for SUSY $n$--point  $\Gamma^{(n)}$  proper vertices type of the 2--point
supercorrelator $C(z_2,z_1)$, Eq.(\ref{53}) and the self--energy
superfunction $\Sigma (z_2,z_1)$. Obviously, under
${\cal D}^{(\alpha)}\equiv \partial_t$,
${\cal D}^{(\alpha)}\equiv {\cal D}_-^{(-)}$
conditions (\ref{C.6}) mean that above--mentioned
supercorrelator depends on   $t_2-t_1$ and
$\overline\theta_2-\overline\theta_1$ differences only:

\begin{eqnarray}
&&C_\varphi(z_2, z_1)
= S(t_2-t_1) +(\overline\theta_2-\overline\theta_1) \Bigl[
G_+(t_2-t_1) \theta_2 - G_-(t_2-t_1) \theta_1 \Bigr]\label{C.7}
 \end{eqnarray}
where the space dependence is suppressed, for brevity. In accordance
with the SUSY field  definition  (\ref{35}), one has:

\begin{eqnarray}
&&S(t_2-t_1) = \left< \eta(t_2) \eta(t_1)  \right>,\nonumber\\
&&G_+(t_2-t_1) = \left<\varphi(t_2)
\eta(t_1) \right> \vartheta(t_1-t_2)=\left<\overline\psi(t_2)
\psi(t_1) \right>\vartheta(t_1-t_2),\label{C.8}\\
&&G_-(t_2-t_1) = \left<\eta(t_2)
\varphi(t_1) \right>\vartheta(t_2-t_1)=\left<\overline\psi(t_1)
\psi(t_2) \right>\vartheta(t_2-t_1)\nonumber \end{eqnarray}
where the step function  $\vartheta(t)=1$ for $t>0$ and
$\vartheta(t)=0$ for $t<0$. By virtue of the causality principle,
the advanced Green function $G_+ (t_2-t_1)$ which is the factor before
$\overline\theta_1\theta_2$, vanishes at $t_1<t_2$, as required. On
the other hand, the retarded Green function $G_-(t_2-t_1)=0$ at
$t_1>t_2$, to be the coefficient of $\overline\theta_2\theta_1$.
Moreover, the symmetry condition $C(z_2,z_1)=C(z_1,z_2)$ gives rise to
the equations $S(t_2-t_1)=S(t_1-t_2)$, $G_-(t_2-t_1)=G_+(t_1-t_2)$.
Inserting the operator $\overline{\cal D}^{(-)}_-$ into the Ward identity
(\ref{C.6}) results in the equation

\begin{eqnarray}
&& 2\dot S(t)= G_+(t)-G_-(t) \label{C.9} \end{eqnarray}
that is  the
fluctuation--dissipation relation.
All the above statements hold for the self--energy function
$\Sigma(z_2,z_1)$ with the components $\Sigma(t_2-t_1)$,
$\Sigma_\pm(t_2-t_1)$ which replace $S(t_2-t_1)$, $G_\pm(t_2-t_1)$,
correspondingly.

It is worthwhile to point out specially the relations in
Eqs.(\ref{C.8}) which connect the Bose correlators for the components
$\eta$, $\varphi$, and the Fermi ones for components
$\psi$, $\overline\psi$.
The above--used Ward identities (\ref{C.6}) allow to obtain these
relations as a trivial consequence of the SUSY field definition
(\ref{35}).  But such equations can be obtained also more simply.
Indeed, the Fermi correlator
$\left<\overline \psi\psi\right>$ is equal to
$\left<(\delta^2 V/\delta\eta^2)^{-1}\right>$ in accordance with Eqs.(\ref{21}),
(\ref{34}). On the other hand, using the susceptibility definition
and Eqs.(\ref{22}), (\ref{26}) we have
$\left<\eta\varphi\right>=\left<\delta\eta/\delta\varphi\right>=
\left<(\delta\varphi/
\delta\eta)^{-1}\right>=\left<(\delta^2 V/\delta\eta^2)^{-1}\right>$
for the Bose correlator Q.E.D.

Obviously, to pass to the correlator of the two--component
field (\ref{11}), it is necessary to replace the factors
$(\overline\theta_2-\overline\theta_1)\theta_2$,
$(\overline\theta_2-\overline\theta_1)\theta_1$ in Eq.(\ref{C.7}) by
the nilpotent coordinates $\vartheta_2$, $-\vartheta_1$,
respectively, and to omit the term with $\overline\vartheta_2\vartheta_1$.
The SUSY correlator $C_f(z_2,z_1)=T_-(z_2)T_-(z_1) C_\varphi(z_2,z_1)$
corresponding to the superfields (\ref{27}) and (\ref{44}), takes the form

\begin{eqnarray}
&& C_f(z_2,z_1)=S + \overline\theta_2\theta_2 m_+
+\overline\theta_1\theta_1 m_- -
\overline\theta_2\theta_1G_-
-\overline\theta_1\theta_2 G_+,  \label{C.10} \end{eqnarray}
where the arguments $t_2-t_1$ are suppressed in the factors $S$, $m_\pm$,
$G_\pm$. Moreover, in view of Eq.(\ref{22}), new functions are
introduced (cf. Eqs.(\ref{C.8}))

\begin{eqnarray}
&& m_+(t) =\left< \eta(-t)\right> \vartheta(-t) f_{\rm ext}, \qquad
m_-(t)=\left< \eta(t) \right>\vartheta(t)f_{\rm ext}, \qquad f_{\rm
ext}\equiv -f \label{C.11} \end{eqnarray}
to represent the connection
between the averaged  order parameter $\left<\eta(t)\right>$ and the
external force $f_{\rm ext}\equiv -f$ (note that the latter is
switched at time $t=0$ and remains constant).  The correlators
(\ref{C.11}) are related to the Green functions as follows:

\begin{eqnarray}
&& G_\pm(t) = m_\pm (t) \pm \dot S(t) \label{C.12} \end{eqnarray}
and possess the symmetry condition $m_+(-t)=m_-(t)$.

{\Large Captures}

\vspace{1cm}

Fig.1 Dependence of the macroscopic memory parameter $q_0$ on the
quenched disorder intensity $h$.

Fig.2 Dependencies of the ergodicity breaking temperature $T_0$
(solid line), and the freezing temperature $T_f$ (thin line) on the
quenched disorder intensity $h$ ($u=0.5$, $v=0$) for different values of the
effective interaction parameter: a) $w$=0.5;  b) $w$=0.2.

Fig.3 Temperature dependencies of the thermodynamic and adiabatic
susceptibilities $\chi$ and $\chi_0$   for:
a) different values of the quenched disorder intensity $h$ (curves 1, 2
correspond to $h=0, 4$) at  $w=0.5$, $u=0.5$, $v=0$;
b) different values of the effective interaction parameter (curves 1, 2
correspond to $w=0.5, 0.2$) at $h=4$, $u=0.5$, $v=0$.

Fig.4 Dependencies of the ergodicity breaking temperature $T_0$
on the quenched disorder intensity $h$ and: a)
effective interaction parameter $w$ at $u=0.5$, $v=0$; b) proper
interaction parameter $v$ at $u=1$, $w=0.5$.

Fig.5 The shift of the temperature dependencies
of the thermodynamic susceptibility $\chi(T)$ caused by variation of:
a) effective interaction parameter $w$ at $u=0.5$, $v=0$ (curves 1,
2, 3 correspond to $w=$0.5, 1, 1.5);
b) proper interaction parameter $v$ at $u=0.5$, $w=0.5$ (curves 1,
2, 3 correspond to $v$=0, 1, 1.5);
c) anharmonicity ratio $u$ at $w=0.5$, $v=0$ (curves 1,
2, 3 correspond to $u=$0.5, 1, 1.5).

Fig.6 Temperature dependencies  ($u=w=0.5$, $v=0$) of:
a) non--ergodicity parameter $\Delta$;
b) microscopic memory parameter $q$.
(curves 1, 2 correspond to $h=$0, 4).

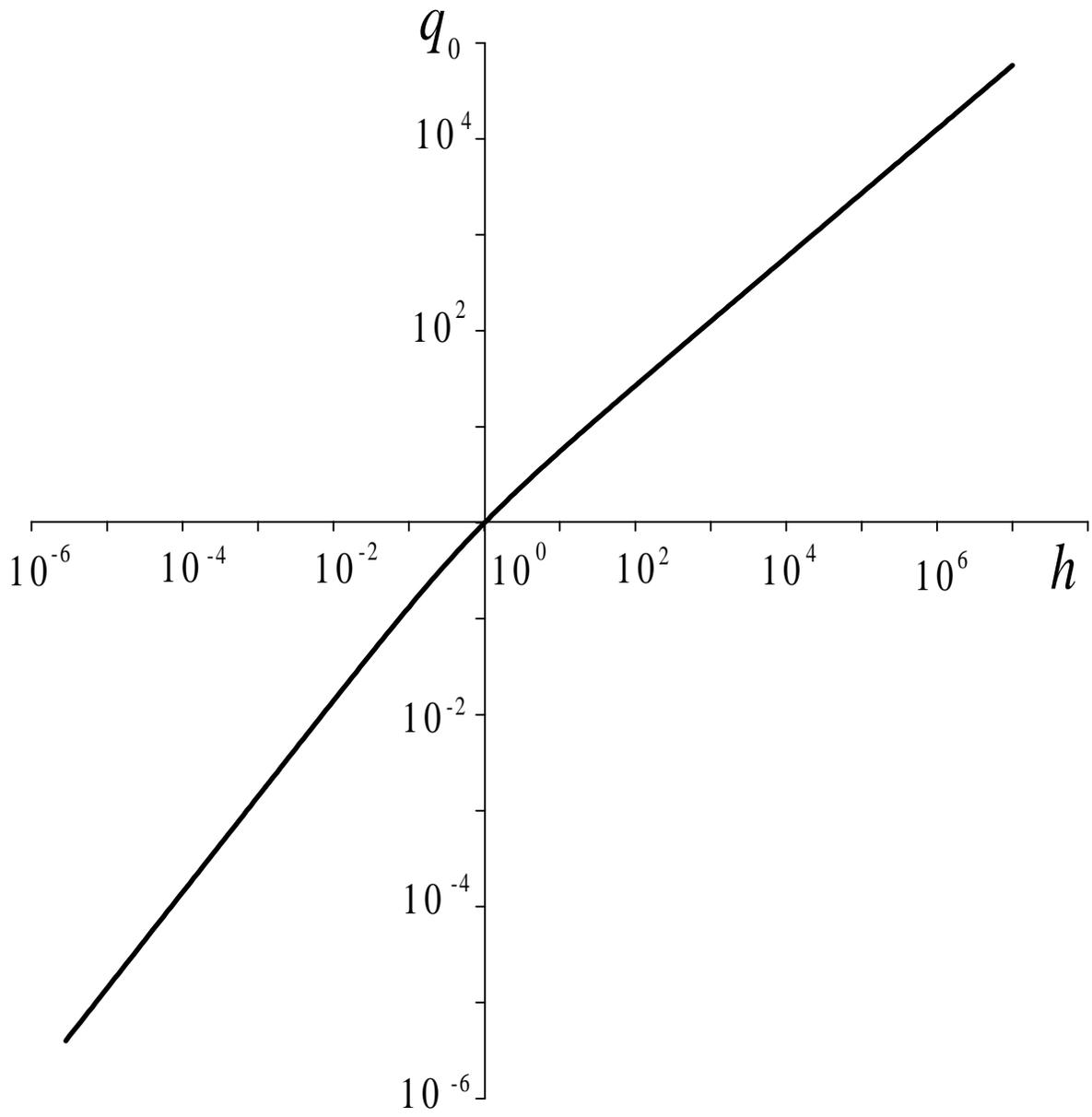
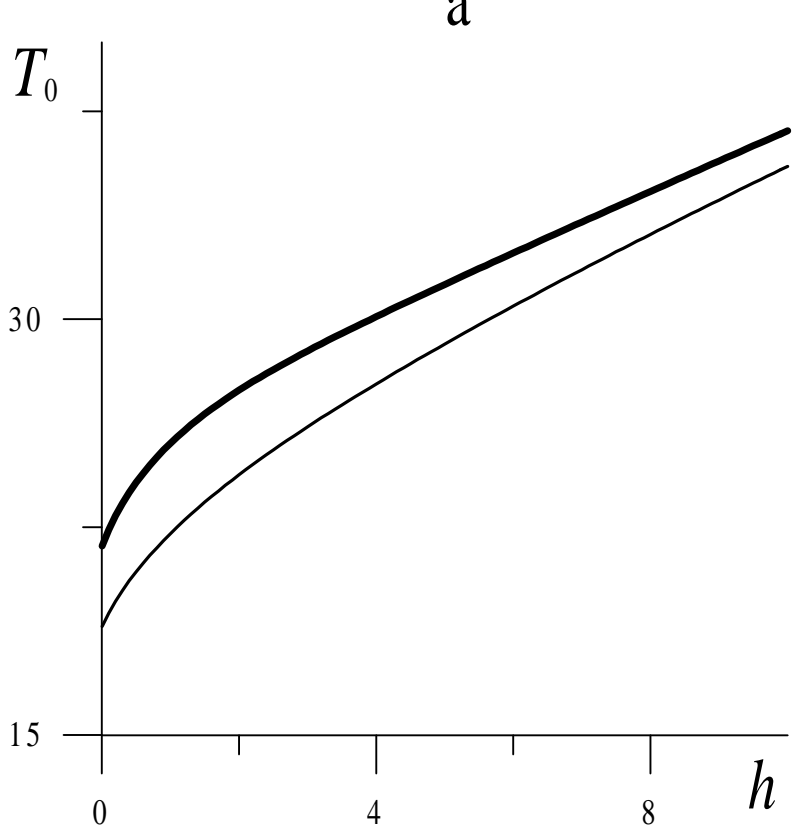
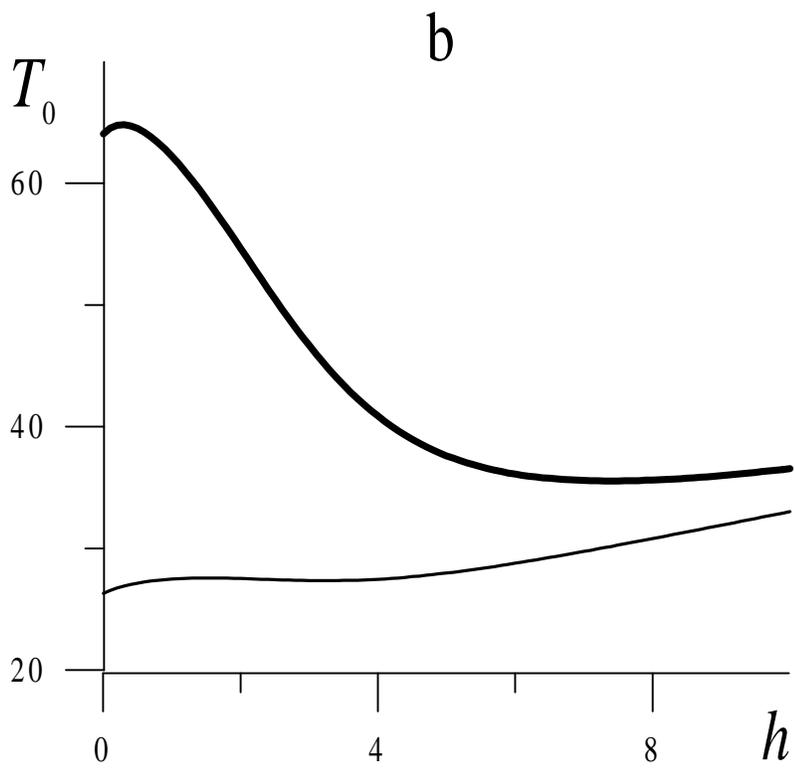
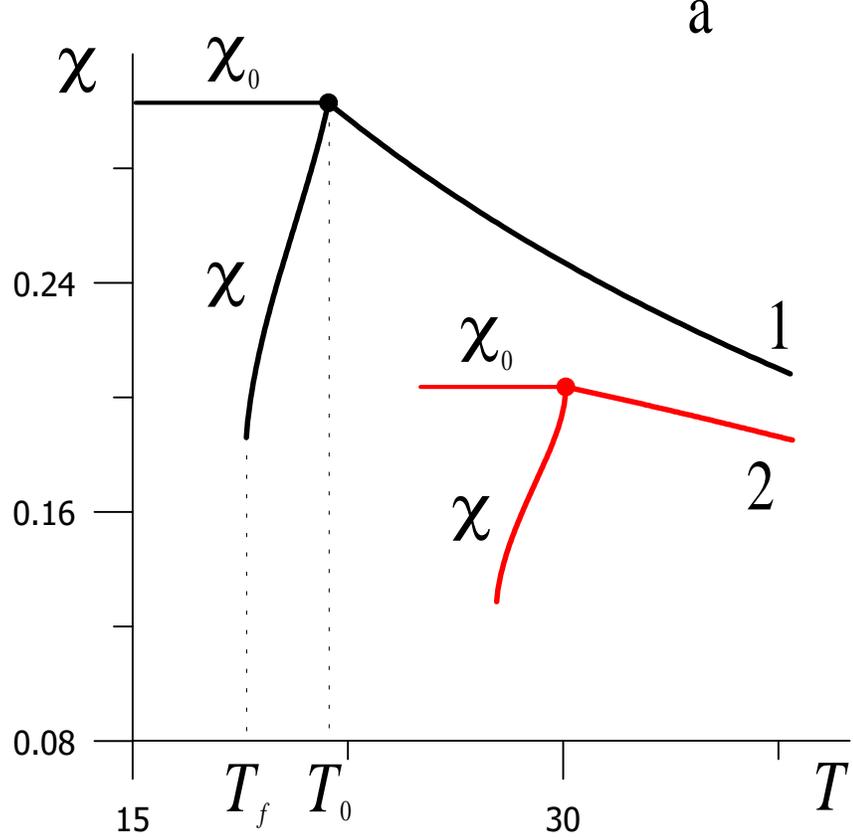
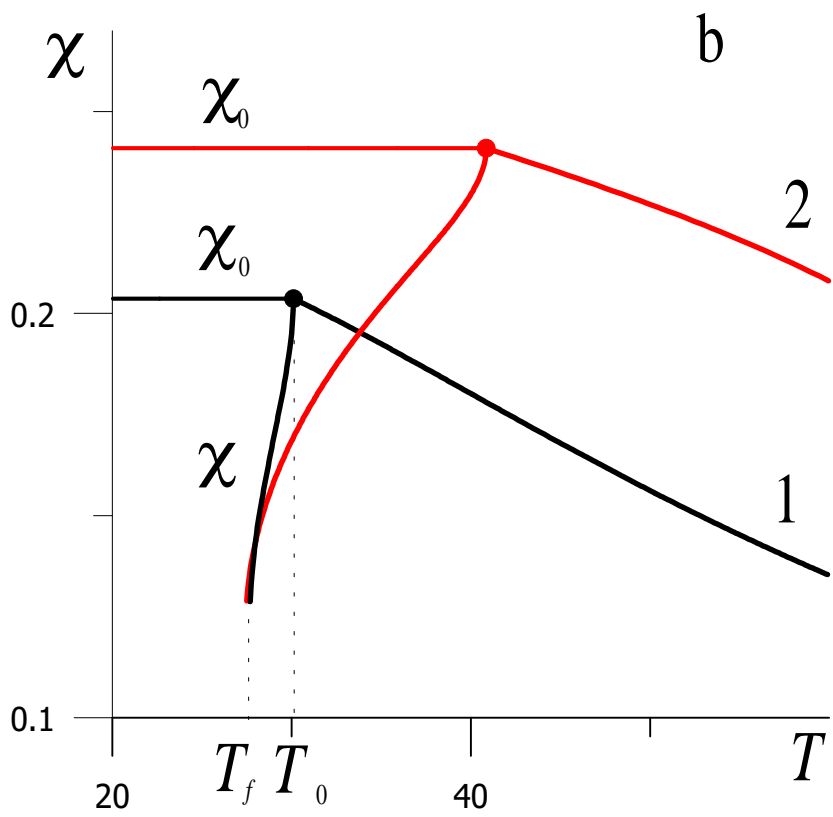
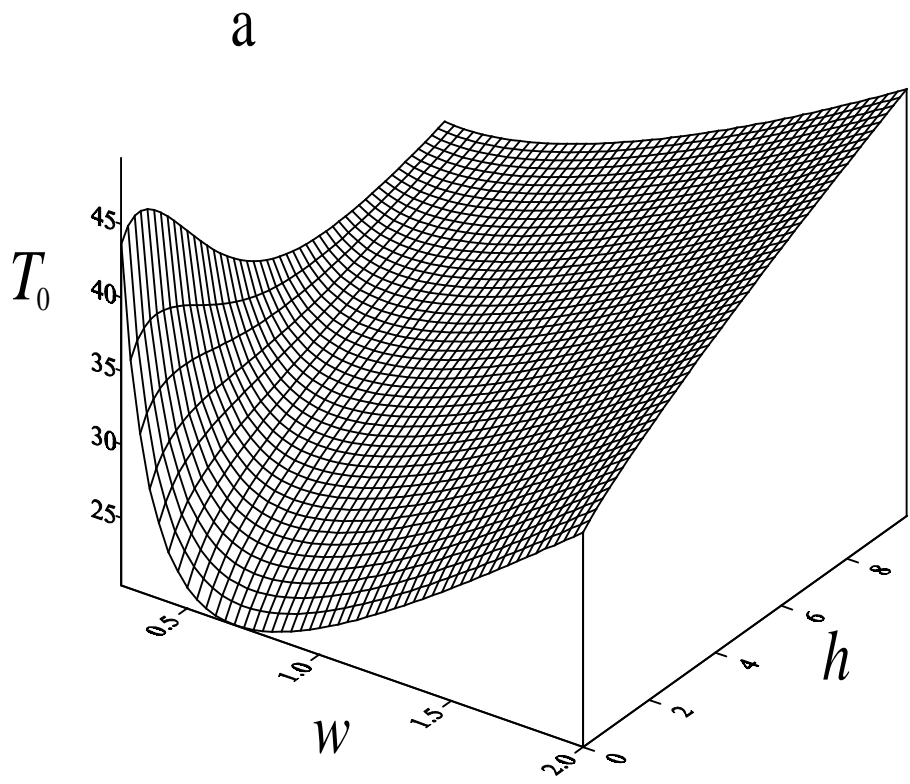
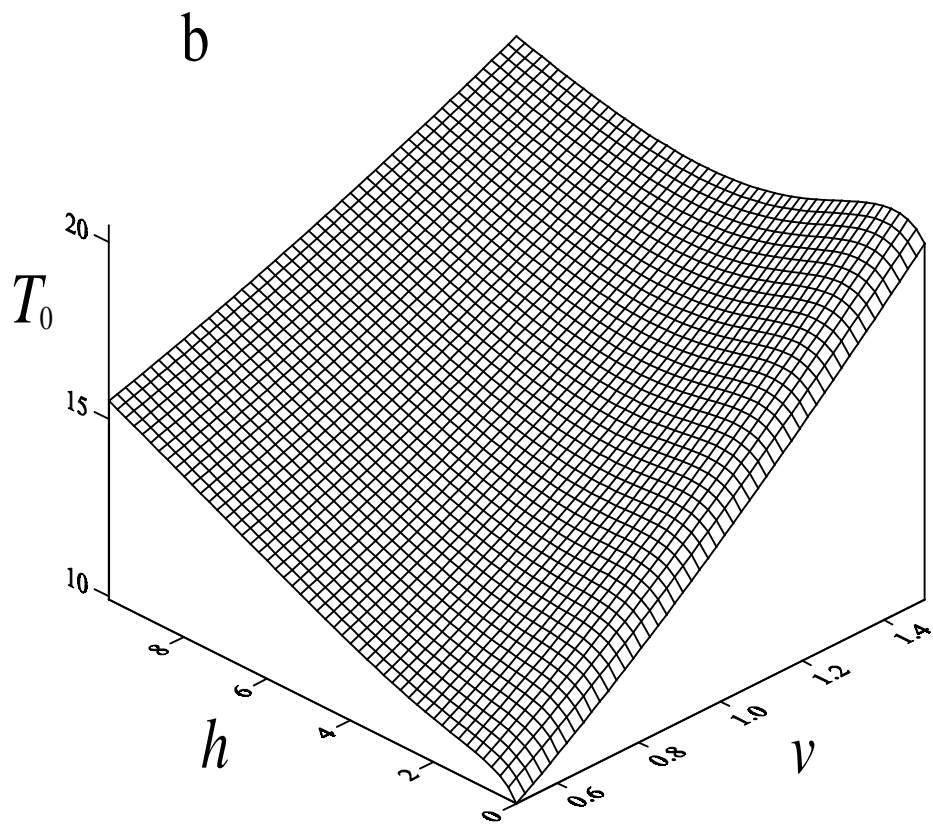
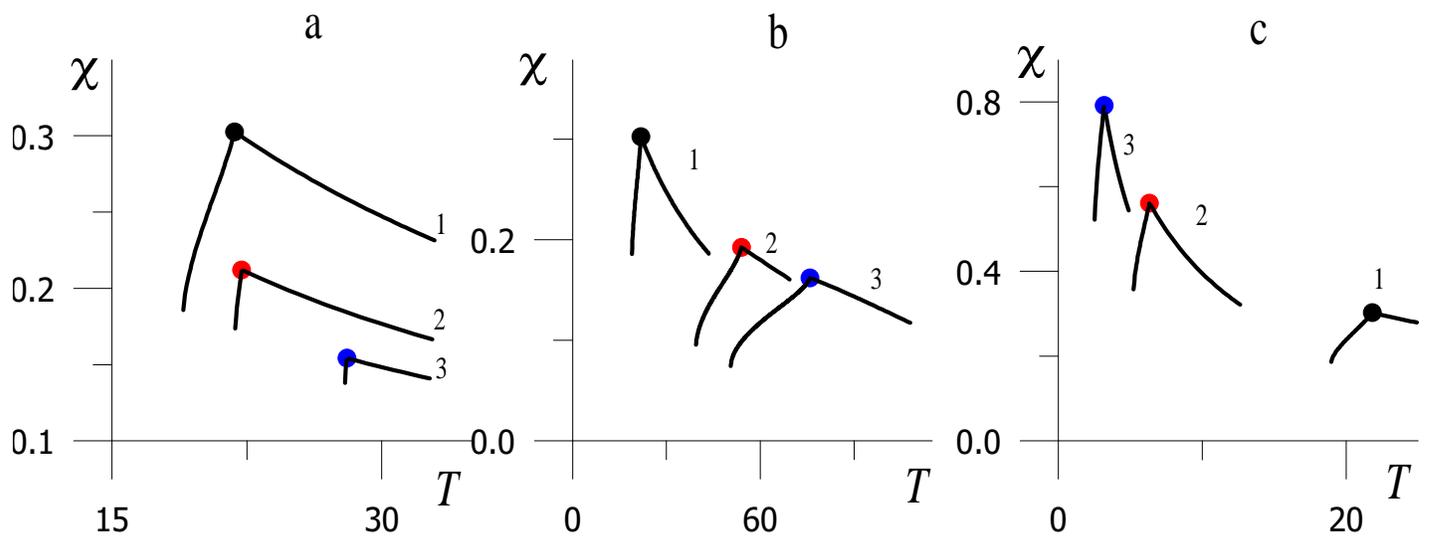
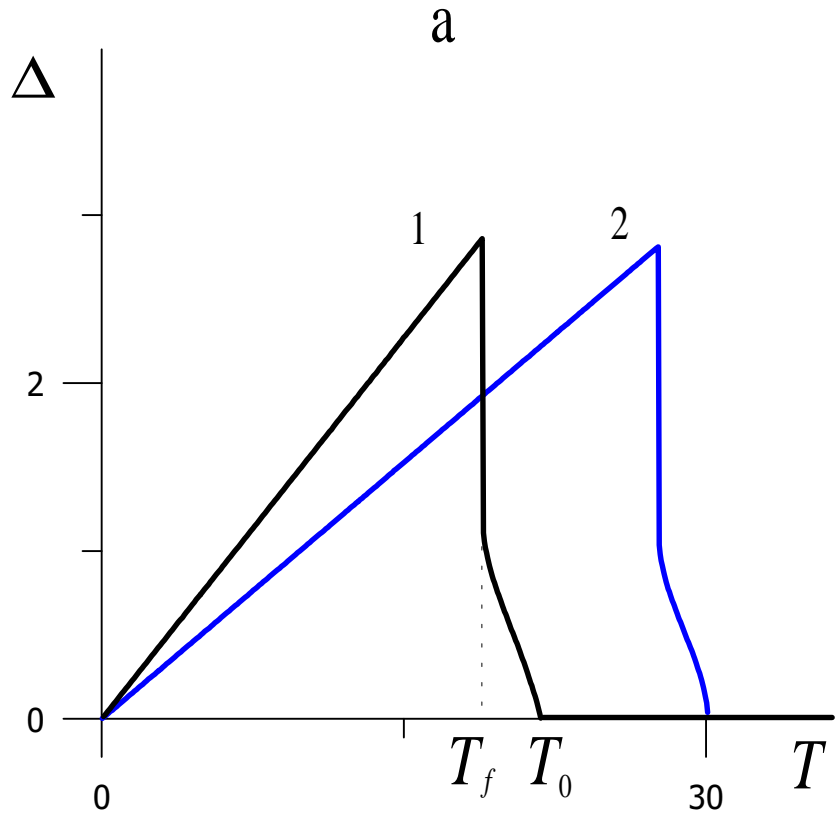
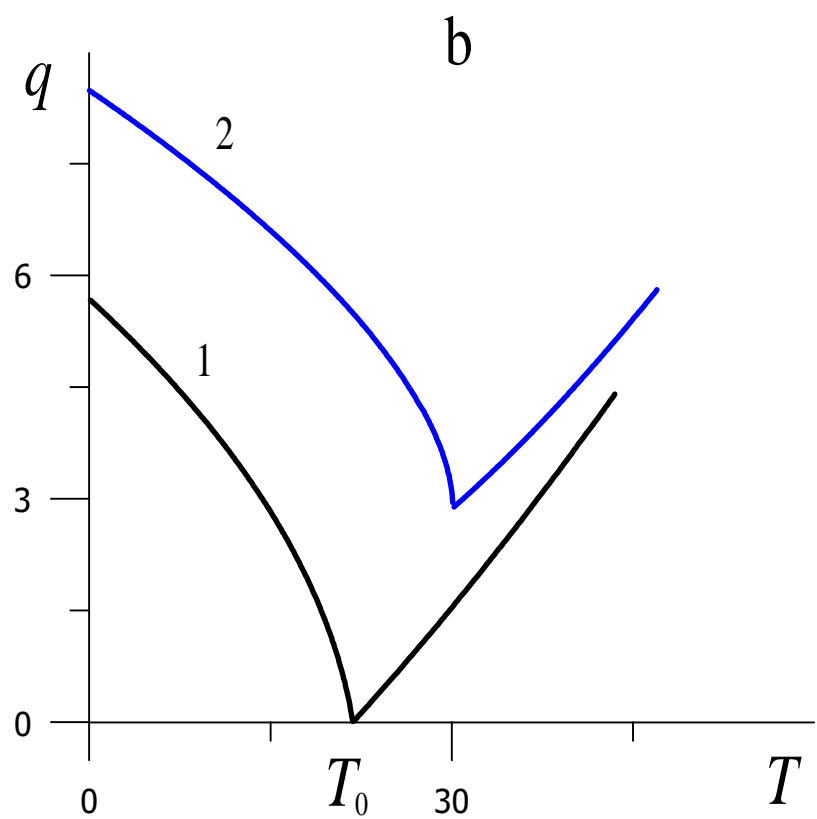
\end{document}